\documentclass[twoside]{article}
\usepackage[a4paper]{geometry}
\usepackage[T1]{fontenc} 
\usepackage{RR}
\usepackage{hyperref}
\usepackage{color}
\usepackage{amsfonts,amssymb}
\usepackage{url}
\usepackage{algpseudocode}
\usepackage{rotating}
\usepackage{multirow}

\usepackage{footnote}
\makesavenoteenv{eqnarray}

\newcommand{\R}{\mathbb{R}}

\newcommand{\PP}{\mathbb{P}}
\newcommand{\ud}{\textrm{d}}

\usepackage[utf8x]{inputenc}
\usepackage[francais]{babel}

\usepackage[usenames,svgnames]{xcolor}

\newcommand{\paulo}[2]{#2}
\newcommand{\roy}[2]{#2}
\RRNo{8072}
\RRdate{September 2012}
\RRauthor{
Shubhabrata Roy
  \and
Thomas Begin
  \and
Patrick Loiseau%
\and Paulo Gonc\c alves
}
\authorhead{Roy \& Begin \& Loiseau \& Gonc\c alves }
\RRetitle{A Versatile  Model for VoD Buzz Workload: Identification, Numerical Validation and Applications in Dynamic Resource Management}
\RRtitle{Un mod\`{e}le de trafic adapté à la volatilité de charge d'un service de vidéo à la demande: Identification, validation et application à la gestion dynamique de ressources.}
\titlehead{VoD Buzz Workload and its application in Dynamic Resource Management}
\RRnote{Work described in this report has been supported by the EU FP7 project SAIL}
\RRresume{La gestion dynamique de ressources est un élément clé du paradigme de {\it cloud computing} et plus récemment de celui de {\ cloud networking}. Dans ce contexte d'infrastructures virtualisées, la réduction des coûts associés à l'utilisation et à la ré-allocation des ressources contraint les opérateurs et les utilisateurs de clouds à une gestion rationnelle de celles-ci. 
Dans ce travail nous proposons une description probabiliste des besoins liée à la volatilité de la charge d'un service de distribution de vidéos à la demande. Cette description peut alors servir de consigne (input) à la provision et à l'allocation dynamique des ressources nécessaires. Notre approche repose sur la construction d'un modèle stochastique inspiré des modèles de Markov standards de propagation épidémiologique, capable de reproduire des variations soudaines et intenses d'activité ({\it buzz}). Nous proposons alors une procédure heuristique d'identification du modèle à partir de séries temporelles du nombre d'utilisateurs connectés au serveur. Les performances d'estimation de chacun des paramètres du modèle sont évaluées numériquement, et nous vérifions l'adéquation du modèle aux données en comparant les distributions des états stationnaires ainsi que les fonctions d'auto-corrélation des processus. 
\newline
Les propriétés markoviennes de notre modèle garantissent qu'il vérifie un principe de grandes déviations permettant de caractériser statistiquement l'ampleur et la durée d'évènements extrêmes et rares tels que ceux produits par les {\it buzzs}.  C'est cette propriété que nous exploitons pour dimensionner le volume de ressources (e.g. bande-passante, nombre de serveurs, taille de buffers) à prévoir pour réaliser un bon compromis entre coût de re-déploiement des infrastructures et qualité de service. Cette approche probabiliste de la gestion des ressources ouvre des perspectives sur les  politiques de  {\it Service Level Agreement} adaptées aux {\it clouds} et servant au mieux les intérêts des opérateurs de réseaux, de services et de leurs clients.
}
\RRabstract{
Dynamic resource management has become an active area of research in the Cloud Computing paradigm. Cost of resources varies significantly depending on configuration for using them. Hence efficient management of resources is of prime interest to both Cloud Providers and Cloud Users. In this report we suggest a probabilistic resource provisioning approach that can be exploited as the input of a dynamic resource management scheme. Using a Video on Demand use case to justify our claims, we propose an analytical model inspired from  standard models developed for epidemiology spreading, to represent sudden and intense workload variations. As an essential step we also derive a heuristic identification procedure to calibrate all the model parameters and evaluate the performance of our estimator on synthetic time series. We show how good can our model fit to real workload traces with respect to the stationary case in terms of steady-state probability and autocorrelation structure. We find that the resulting model verifies a Large Deviation Principle that  statistically characterizes extreme rare events, such as the ones produced by ``buzz effects" that may cause workload overflow in the VoD context.\newline
This analysis provides valuable insight on expectable abnormal behaviors of  systems. We exploit the information obtained using the Large Deviation Principle for the proposed Video on Demand use-case for defining policies (Service Level Agreements). We believe these policies for elastic resource provisioning and usage may be of some interest to all stakeholders in the emerging context of cloud networking.}
\RRmotcle{R\'{e}seaux, Cloud, Gestion probabiliste des Ressources, Modèles Epidémiques , Générateur de Charge, Estimation Statistique, Principe de Grandes Déviations, {\it Service Level Agreement}, Vidéo à la Demande, Buzz}
\RRkeyword{Cloud Networking,Probabilistic Resource Management, Epidemic Model, Workload Generator, Statistical Estimation, Large Deviation Principle, Service Level Agreements, Video on Demand, Buzz}
\RRprojet{RESO}
 \RCGrenoble 

\begin{document}
\makeRR   
\tableofcontents
\section{Introduction}
\label{sec:intro}

\begin{figure}[h]
\centering
\hspace*{-3mm}
\begin{tabular}{cc}
\begin{turn}{90}{\hspace*{5mm} Number of downloads / day} \end{turn} &
\hspace*{-4.5mm}\includegraphics[width=0.7\columnwidth]{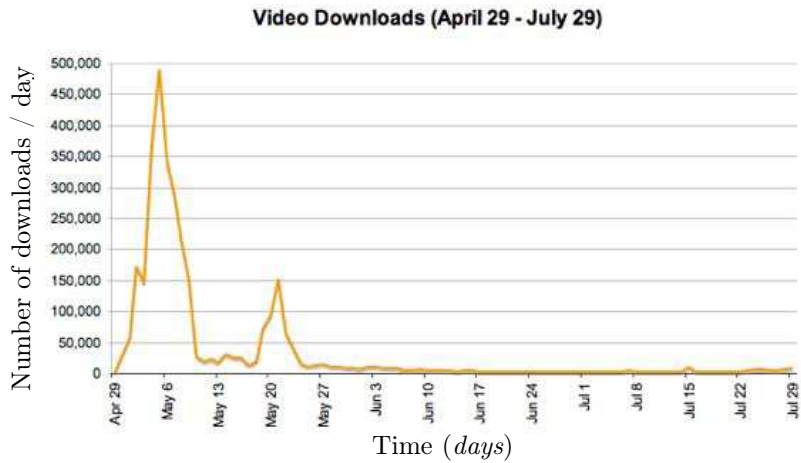}\\[-2mm]
& Time ({\em days})\\
\end{tabular}
\caption{\small Video server workload: time series displaying a characteristic pattern of flash crowd (buzz effect). Trace obtained from \cite{website:waxy}.}
\label{fig:viral}
\end{figure}

In recent trend of data-intensive applications with pay-as-you-go execution in a cloud environment, there are new challenges in system management and design to optimize the resource utilization. Types of the application, deployed in a cloud, can be very diverse. \roy{There are some applications that need to be rapidly cloned or re-allocated, like a pre-production environment.}{} Some applications exhibit highly varying demand in resources. In this paper we consider a Video on Demand (VoD) system as a relevant example of a data-intensive application where bandwidth usage varies rapidly over time. \newline
A VoD service delivers video contents to consumers on request. According to Internet usage trends, users are increasingly getting more involved in the VoD and this enthusiasm is likely to grow. According to 2010 statistics a popular VoD provider like Netflix accounts for around 30 percent of  the peak downstream traffic in the North America and is the ``largest source of Internet traffic overall" \cite{website:sandvine}. Since VoD has stringent streaming rate requirements, each VoD provider needs to reserve a sufficient amount of server outgoing bandwidth to sustain continuous media delivery (we are not considering IP multicast here). However, resource reservation is very challenging in a situation, when a video becomes popular very quickly leading to a \emph{flood} of user requests on the VoD servers. This situation, also known as a ``buzz", demands an adaptive resource allocation strategy to cope with the sudden (and significant) variation of workload. Following is one example of ``buzz" (see Figure~\ref{fig:viral}) where interest over a video ``Star Wars Kid" \cite{website:waxy} grew very quickly within a short timespan. According to \cite{website:bbc} it was viewed more than 900 millions times within a short interval of time making it one of the top viral videos.
Such bandwidth volatility creates significant challenges to meet, namely, both the desired QoS and efficient resource allocation. \roy{}{A sensible} approach to this problem is to help the providers in better understanding and capturing the underlying characteristics of their applications. For example, if the information diffusion process follows a gossip\roy{}{-} (or epidemic\roy{}{-}) behavior, the rate at which a \roy{customer}{viewer} \roy{}{keeps gossiping about a video and for how long (in average)}.\newline
In this report we follow a constructive approach to propose a stochastic epidemic workload generator for a VoD system based on a Markov model. We show that it succeeds to reproduce the traffic volatility, as exhibited in a real trace, including the buzz occurrence. But the principal interest of our model is that it verifies a Large Deviation Principle (LDP) that gives a probabilistic description of the mean workload of the system over different time scales. It thus adequately allows for statistically characterizing extreme rare events such as the ones produced by buzz transients. Our ultimate objective is to exploit this large deviation information as an input of an \roy{adaptive}{probabilistic} resource management scheme. However, in order the proposed model to conform with this objective, it needs to be ``identifiable" and easily calibrated on real data. The corresponding estimation procedure may not be trivial, since the VoD model is a non-parsimonious model and accounts for \roy{a}{}complex dynamics. In this report we propose a complete framework for the operators to identify the VoD model parameters based on a server workload trace\newline
After parameter estimation we devise two possible and generic ways to exploit the large deviation information in the context of probabilistic resource provisioning. They can serve as the input of resource management functionalities of the Cloud environment. It is evident that we can not define elasticity without the notion of a time scale; the Large Deviation Principle (LDP) is capable of automatically integrating the time resolution in automatic description of the system. It is to be noted that Markovian processes do satisfy the LDP, but so do some other models as well. Hence, our proposed probabilistic approach is very generic and can adapt to address any provisioning issues, provided the resource volatility can be resiliently represented by a stochastic process for which the LDP holds true. \newline
In a nutshell our contributions in this report include: 
\begin{itemize}
\item A Markov based versatile model to generate VoD workload, 
\item A heuristic identification procedure for the proposed workload model, 
\item A numerical evaluation of the estimator \roy{of}{for} each parameter\roy{s}{} of the model,
\item A real case study to assess the adequacy of our model to fit video workload traces,
\item An analysis of the Large Deviation property of the proposed Markovian model,
\item A discussion on the generic ways to exploit the large deviation information in the context of probabilistic resource provisioning.
\end{itemize}
Moreover, since we followed a constructive approach, each parameter of the model accounts for a specific component of the system, and so, its estimated value also permits to quantify the importance of the corresponding dynamic effect.\newline
Rest of the paper is organized as follows. In Section \ref{sec:related} we discuss the related works. We describe our model and further analyze it in Section \ref{sec:vod}. Section \ref{sec:estimation} outlines the parameter estimation procedure and validates the procedure against synthetic workload traces. In Section \ref{sec:validation} we validate both our model and the estimation procedure against the real workload traces. Section \ref{sec:ldp} presents Large Deviation Principle and numerical interpretations of the Large Deviation Spectrum. Section \ref{sec:management} deals with the probabilistic provisioning scheme, derived from the Large Deviation Spectrum for our use case. Finally we conclude and discuss future works in Section \ref{sec:conclusion}.\newpage
\section{Related Work}
\label{sec:related}
Information dissemination in a  VoD system has been an active area of research. In \cite{EugEpi2004}, it has been already demonstrated that the epidemic algorithms can be used as an effective solution for information dissemination in a VoD like P2P systems. \roy{}{However, in this model an individual process must have a precise idea about the total number of processes in the system. Scalability is also another challenge that the authors addressed in this work}. The authors of \cite{BonalEpistream2008} studied random epidemic strategies like the random peer, latest useful chunk algorithm to achieve optimal information dissemination. But main objective of this work is to demonstrate ways to achieve performance trade-offs using unstructured, epidemic live streaming systems. However, it does not bring any information about the underlying dynamics of the \roy{}{streaming} system. Authors of \cite{BakhshiCQST11} similarly discussed an analytical framework for gossip protocols based on the pairwise information exchange between interacting nodes. However, this model only provides an analysis of networks with lossy channels. Another relevant work to our study is derived in \cite{CarPatMatch2010} where the authors proposed an approach to predict workload for \paulo{the}{} cloud clients. They considered an auto-scaling approach for resource provisioning and validated the result with real-world cloud client application traces. However, this work depends on similar past occurrences of the current short-term workload history and is not appropriate to deal with sudden and short large variations of workload, as the ones produced by buzz effects.
Authors of \cite{GarSim2007} show a statistical study of streaming traffic. They analyzed VoD session characteristics, amount and types of media delivered, popularity and daily access profile in order to develop a workload generator. However, the model does not involve the dynamics of the process itself, ergo it is not naturally adapted to infer dynamic resource allocation strategies. Authors of \cite{LiPerf1996}, \cite{MelMul2009} and \cite{Kanrar2012} also develop \roy{a}{}user activity model\roy{}{s} to describe the usage of system resources. \roy{}{Limitation of these models are that they only give} average results. \roy{}{However, dealing with mean workloads might not be sufficient to clearly describe applications because of their potential volatility}. In \cite{CasaleIEEECloud11} authors proposed a maximum likelihood method for fitting a Markov arrival process (MAP) to the web traffic measurements, collected in commonly available HTTP web server traces. This method achieves reasonable accuracy in predictive models for web workloads but lacks intuitive nature to describe users behavior like a gossip based method. In \cite{NiuNOSS11} the authors statistically model traffic volatility in large scale VoD systems using GARCH (generalized autoregressive conditional heteroscedasticity) process. Amazon Cloud-Watch follows this approach and provides a free resource monitoring service to Amazon Web Service customers for a given frequency. Based on such estimates of future demand, each VoD provider can individually reserve a sufficient amount of bandwidth to satisfy {\it in average} its random future demand within a reasonable confidence. However, according to the authors, this technique only models and forecasts the mean demand, or the expected demand whereas the real demand might vary around this predicted mean.\roy{}{ They suggested to provision an additional ``risk premium" to the service providers for tolerating the demand fluctuation}. In another workload model the authors of \cite{Perez-Palacin12} \cite{Gusella91} proposed a Markov Modulated Poisson Process (MMPP) based approach for buzz modeling and then parameter estimation using the index of dispersion. However, the MMPP model includes only short-term memory in the system and the obtained statistics is not \roy{}{physically interpretable to draw inference about the system dynamics}. \newline
The model we derive in section \ref{sec:vod} of this report has the following advantages:
\begin{itemize}
\item It follows a constructive approach, based on a Markov model, 
\item It is identifiable and succeeds to capture workload vo\-la\-ti\-li\-ty, 
\item It satisfies the large deviation properties, that can be exploited to frame dynamic resource allocation strategies.
\end{itemize}
\roy{We believe this model along with its identification procedure can be of some interest to the VoD operators.}{}
\newpage

\section{A VoD system and its modeling}
\label{sec:vod}
\begin{figure}[h]
\centering
\includegraphics[width=0.3\columnwidth]{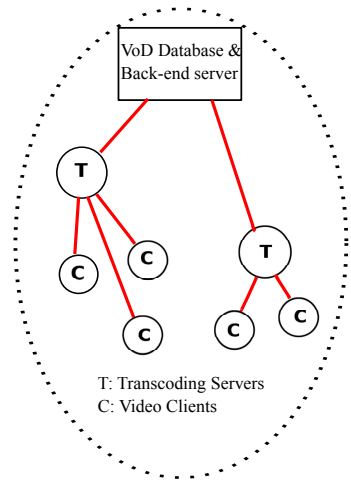}
\caption{\small Basic schematics of a VoD system with transcoding/caching servers}
\label{fig:VoD}
\end{figure}

A VoD service delivers video contents to consumers on request. According to Internet usage trends, users are increasingly getting more involved in the VoD and this enthusiasm is likely to grow. A popular VoD provider like Netflix accounts for around 30 percent of  the peak downstream traffic in the North America and is the ``largest source of Internet traffic overall" \cite{website:sandvine}. In a VoD system, consumers are video clients who are connected to a \textit{Network Provider}. The source video content is managed and distributed by a \textit{Service Provider} from a central data centre. With the evolution of Cloud Computing {and Networking}, the service in a VoD system can be made more scalable by dynamically distributing the caching/transcoding servers across the network providers. Video service providers interact with the network service providers and describe the virtual infrastructures required to implement the service (like the number of servers required, their placements and clustering of resources). The resource provider reserves resource for certain time period and may change it dynamically depending on resource requirement. Such a dynamic approach brings benefits of cost saving in the system through dynamic resource provisioning which is important for service providers as VoD workload is highly variable by nature. However, since {the virtual resources used by Cloud Networking}  have a set-up time which is not negligible, analysis and provisioning of such a system can be very critical from the operators perspective ({\sc capex} versus {\sc opex} trade-off). Figure \ref{fig:VoD} shows a VoD schematic where the back-end server is connected to the data centre and the transcoding (caching) servers are placed across the network providers.
Since VoD has stringent streaming rate requirements, each VoD provider needs to reserve a sufficient amount of server outgoing bandwidth to sustain continuous media delivery. When multiple VoD providers (such as Netflix) are on board to use cloud services from cloud providers, there will be a market between VoD providers and cloud providers, and commodities to be traded in such a market consist of bandwidth reservations, so that VoD streaming performance can be guaranteed.

As a buyer in such a market, each VoD provider can periodically make reservations for bandwidth capacity to satisfy its random future demand. A simple way to achieve this is to estimate expectation and variance of its future demand using historical demand information, which can easily be obtained from cloud monitoring services. As an example, Amazon Cloud-Watch provides a free resource monitoring service to Amazon Web Service customers for a given frequency. Based on such estimates of future demand, each VoD provider can individually reserve a sufficient amount of bandwidth to satisfy {\it in average} its random future demand within a reasonable confidence. However, this information is not helpful in case of a ``buzz" or a ``flash crowd" when a video becomes popular very quickly leading to a \emph{flood} of user requests on the VoD servers.
In situations like the one described in Figure \ref{fig:viral}, variance estimation or more generally steady state distribution can not explain burstiness of such event as time resolution is excluded from the description. The LDP, by virtue of its multi-resolution extension of the classical steady-state distribution, can describe the dynamics of rare events like this, which we believe can be of some interest for the VoD service providers.
\subsection{Markov Model to describe the VoD user behavior}
\label{sec:markov}
{Epidemic models commonly subdivide a population into several compartments:} susceptible (noted $S$) to designate the persons who can get infected, and contagious (noted $C$) for the persons who have contracted the disease. This contagious class can further be categorized into two parts: the infected subclass ($I$) corresponding to the persons who are currently suffering from the disease and can spread it, and the recovered class ($R$) for those who got cured and do not spread the disease anymore \cite{BarDynaPro2008}. There can be more categories {that fall outside} the scope  of our current work.  In these models $(N_{S}(t))_{t \geq 0}$, $(N_{I}(t))_{t \geq 0}$ and $(N_{R}(t))_{t \geq 0}$ are stochastic processes representing the {time} evolution of susceptible, infected and recovered populations respectively.
\newline {Similarly,} information dissemination in a social {network} can be viewed as an epidemic spreading {(through gossip)}, where the ``buzz"  is a special event where interest {for} some {particular} information increases {drastically} within a very short period {of time}. Following {the lines of} related works, we claim that the above mentioned epidemic models can appropriately be adapted to represent the way information spreads among the users in a VoD system.  In the case of a VoD system, infected $I$ refers to the people who  are currently watching the video and can spread the information about it. {In our setting,} $I$ directly represents the current workload which is the current aggregated video requests from the users. Here, we consider the workload as the total number of current viewers, but it can also refer to total bandwidth requested at the moment. The class $R$  refers to the past viewers. {In contrast to the classical epidemic case, we introduce a memory effect in our model, assuming that the $R$ compartment can still propagate the gossip during a certain random latency period.}
{Then, we define the probability within a small time interval ${\rm d}t$, for a susceptible individual to turn into an active viewer, as follows}:
\begin{equation}
\mathbb{P}_{S\to C} = {(l+ (N_I(t)+N_R(t)) \, \beta) {\rm d}t} + o({\rm d}t)
\label{eq:transition-prob}
\end{equation}
where $\beta > 0$ is the rate of information dissemination per unit time and $l>0$ fixes the rate of spontaneous viewers.
The  instantaneous rate of newly active viewers in the system at time $t$ is thus:
\begin{equation}
\lambda(t) = l+ (N_I(t)+N_R(t))\beta.
\label{eq:infinity}
\end{equation}
Equation (\ref{eq:infinity}) corresponds to the arrival rate $\lambda(t)$ of a non-homogeneous (state dependant) Poisson process. This rate varies linearly with $N_I(t)$ and $N_R(t)$.
%

\begin{figure}[h]
\centering
\includegraphics[width=0.4\columnwidth]{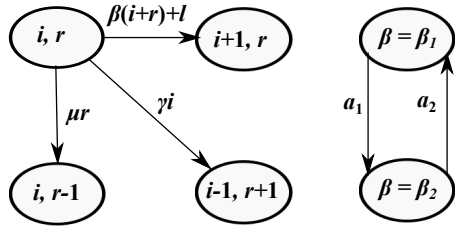}
\caption{\small Markov chain diagram representing the evolution of the Current viewers ($i$) and Past Viewers ($r$) populations with a Hidden Markov Model. }
\label{fig:Markov}
\end{figure}

To complete our model we assume that the watch time of a video is exponentially distributed with rate $\gamma$. As already mentioned, it also deems reasonable to consider that a past viewer will not keep propagating the gossip about a video indefinitely, but  remains active only for a latency random period that we also assume exponentially distributed  with rate $\mu$ (in general $\mu \ll \gamma$). Another important consideration of the model is the maximum allowable viewers ($I_{\textrm{max}}$) at any instant of time. This assumption conforms to the fact that the resources in the system are physically limited. For the sake of numerical tractability and without loss of generality, we also assume the number of past (but spreading rumour) viewers at a given instant to be bounded by a maximum value ($R_{\textrm{max}}$). With these assumptions, and posing ($N_I(t)=i,N_R(t)=r)$ the current state of the Markov processes, the probability that the process reaches a different state $(i' < I_{\textrm{max}},r'< R_{\textrm{max}})$ at time $t+{\rm d}t$  (${\rm d}t$ being small) reads:
\begin{eqnarray}
\lefteqn{\mathbb{P}(i', r' | i, r) } \\
&=& (l+(i+r) \beta){\rm d}t + o({\rm d}t) \hspace*{5mm} \mbox{for } (i'=i+1,r'=r), \nonumber \footnotemark\footnotetext{In a closed system, where the total number of viewers (susceptible, current and past) is constant, say $N$, the transition probability for ($i'=i$+1,$r'=r$) needs to be modified, since it would then depend on the number of susceptible viewers as well,, i.e ($N-i-r$). The transition probability in this case would be $(l+(i+r) {\beta \over N})(N-i-r){\rm d}t + o({\rm d}t)$. Therefore, Eq. \ref{eq:i_mean} and \ref{eq:stability} need to be modified accordingly.}\\
&=& (\gamma i){\rm d}t + o({\rm d}t)  \hspace*{12mm}\mbox{for } (r'=r+1,i'=i-1), \nonumber \\
&=& (\mu r){\rm d}t + o({\rm d}t) \hspace*{17mm} \mbox{for } (r'=r-1,i'=i), \nonumber \\
&=& o({\rm d}t)  \hspace*{45mm} \mbox{otherwise.} \nonumber
 \label{eq:MC2-transition}
\end{eqnarray}
This process defining the evolution of the current viewer and past viewer populations  is a finite and irreducible Markov chain. It is to be noted that $l  > 0$ precludes the process to reach an absorbing state. This chain is ergodic and admits a stationary regime. \newline
Above mentioned descriptions define the mechanism of information dissemination in the community in normal situations. A buzz event differs from this situation by a sudden increase of the dissemination rate $\beta$. In order to adapt the model to buzz we resort to Hidden Markov Model (HMM) to be able to reproduce the change in $\beta$. Without loss of generality we consider only two states. One with dissemination rate $\beta=\beta_{1}$ corresponds to the buzz-free case described above, and another hidden state corresponding to the buzz situation, where the value of $\beta$ increases significantly and takes on a value  $\beta_{2} \gg \beta_{1}$. Transitions between these two hidden and memoryless Markov states occur  with rates $a_{1}$ and $a_{2}$ respectively (see Figure \ref{fig:Markov}). These rates characterize the buzz in terms of frequency, magnitude and duration. 

\begin{figure}[t]
\centering
\begin{tabular}{ccc}
\hspace*{-8mm}case (a) & case (b) & case (c) \\
\hspace*{-8mm}\includegraphics[width=0.40\columnwidth]{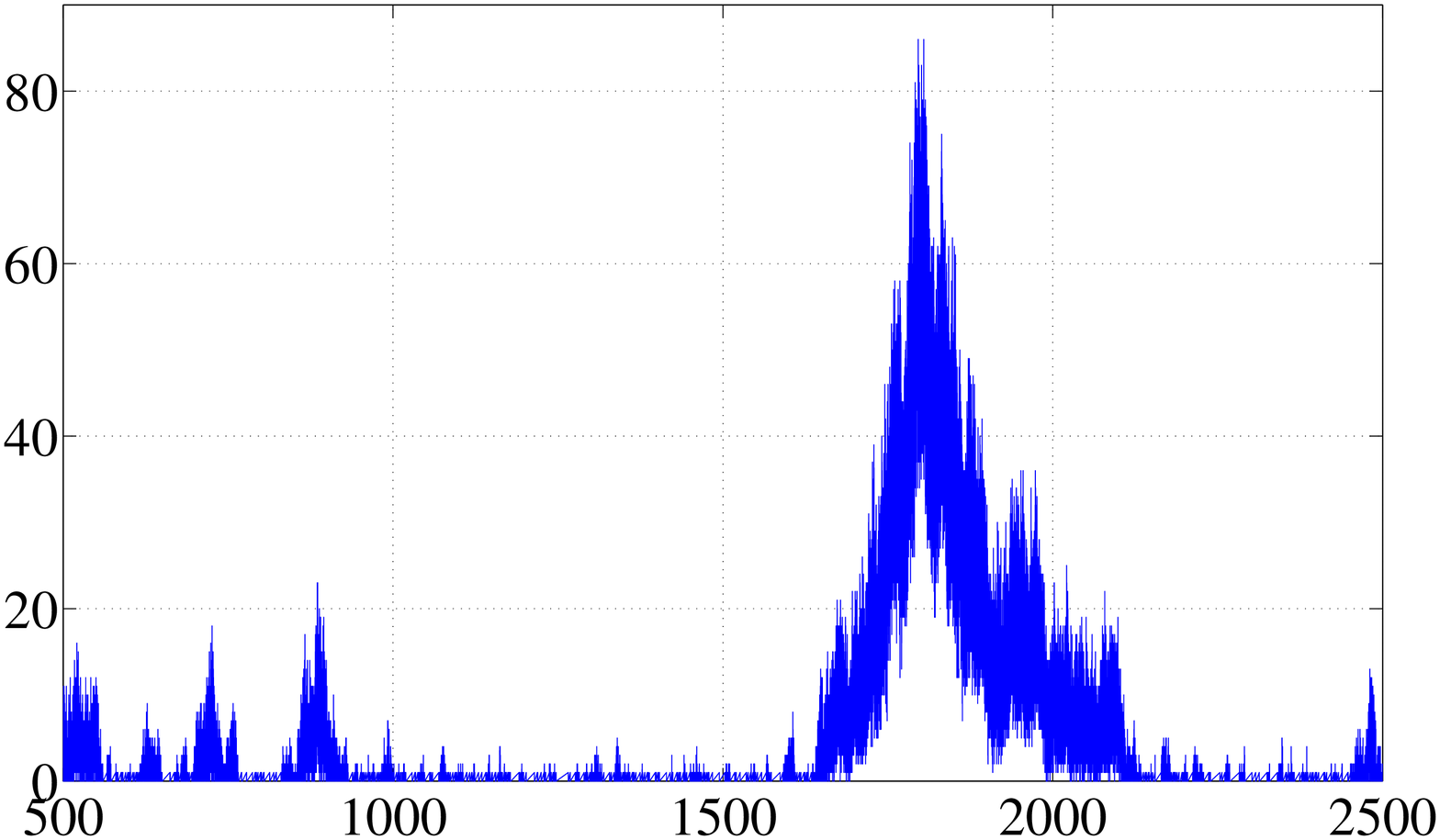} &
\hspace*{-8mm}\includegraphics[width=0.40\columnwidth]{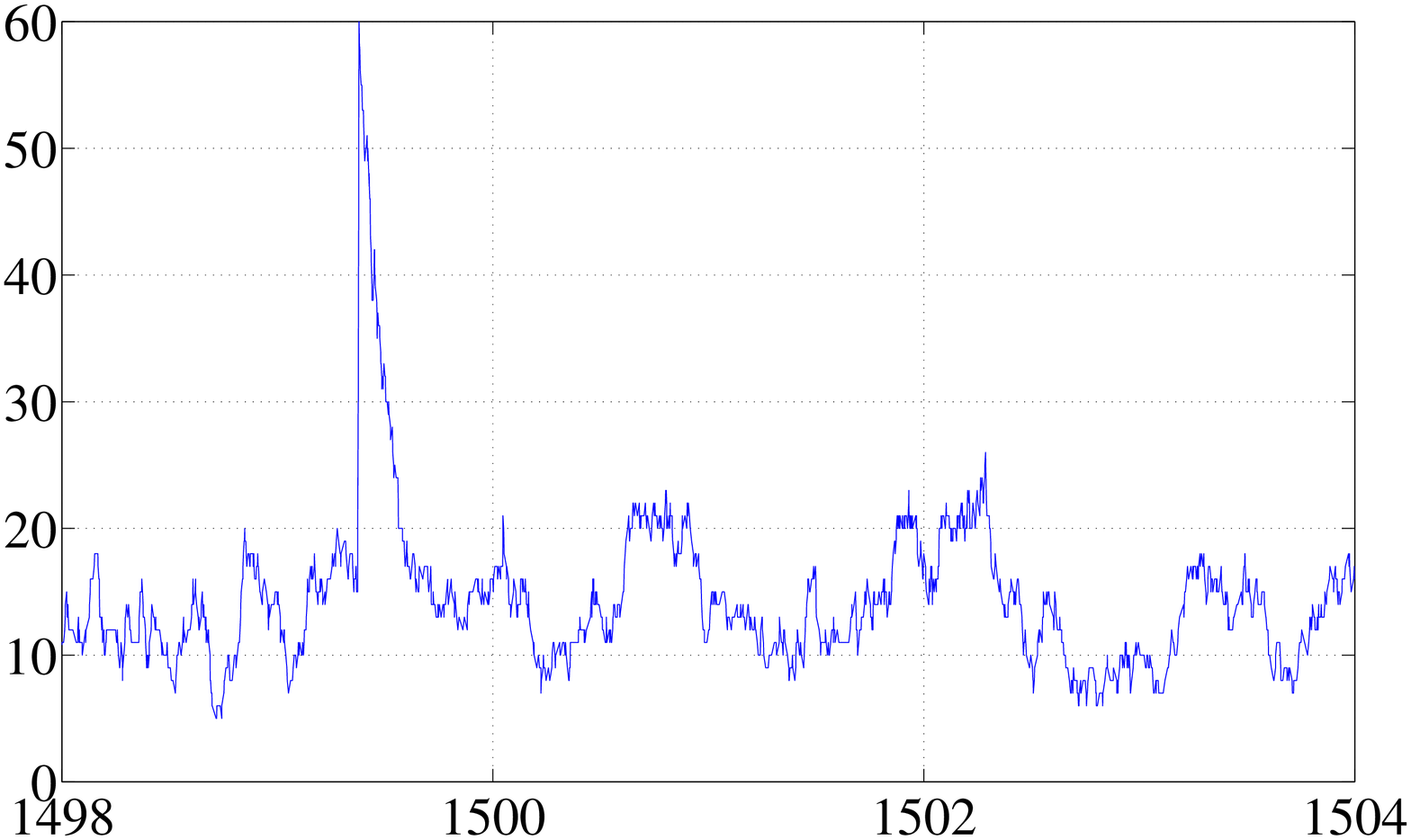} &
\hspace*{-8mm}\includegraphics[width=0.40\columnwidth]{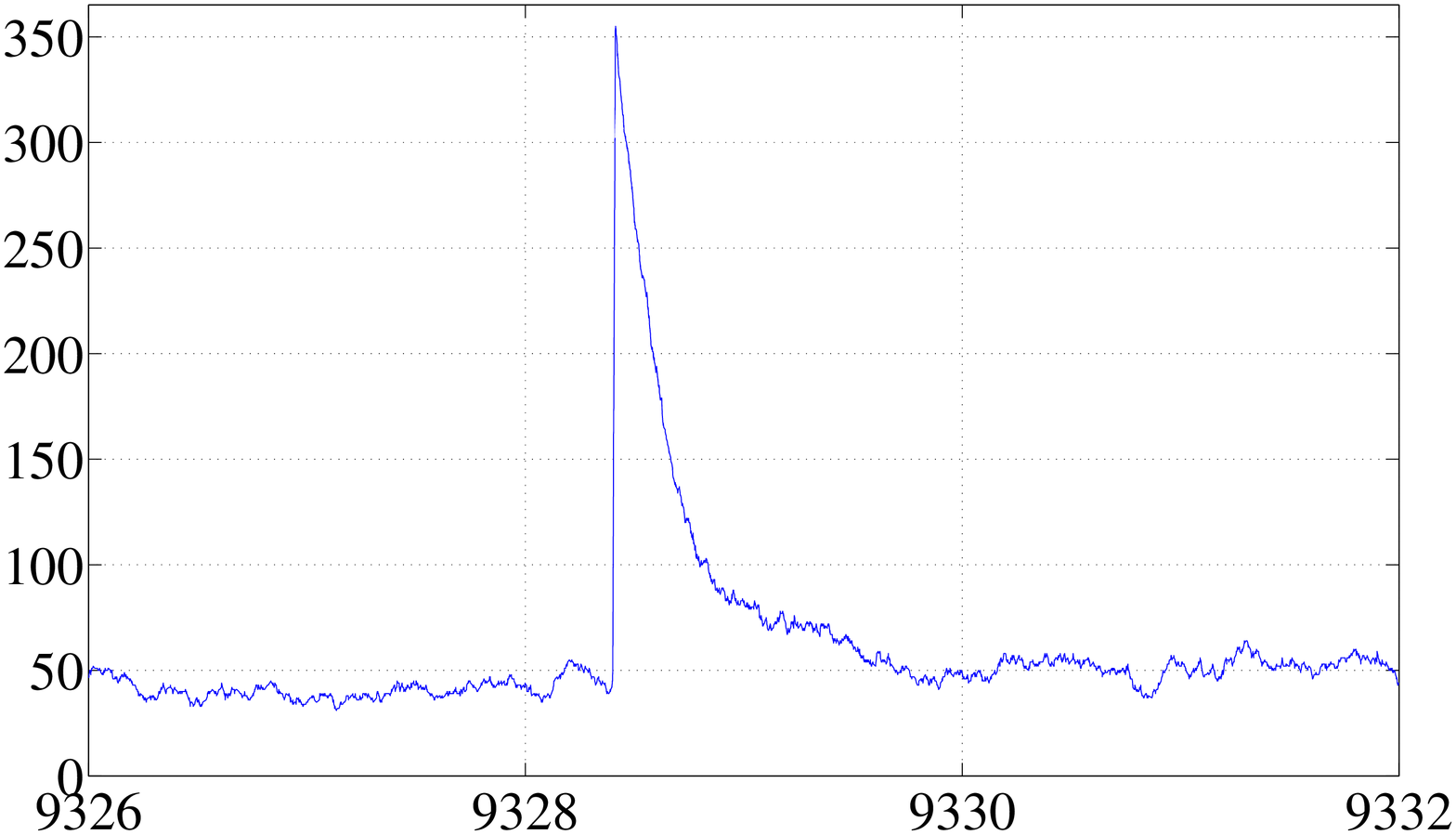}
\end{tabular}
\caption{\small Illustration of our model ability at generating different dynamics of workload $I(t)$. See Table \ref{table1} for the parameter values corresponding to each of these three cases. The $X-$axis corresponds to time (in hours unit) while the $Y-$axis indicates the number of active viewers.}
\label{fig:traces}
\end{figure}

\paulo{}{A closed-form expression for the steady state distribution of the workload ($i$) of this model \roy{}{is not trivial to derive}. However, we could easily express the analytic mean workload of the system solving the flow balance equation, i.e. equaling the incoming and outgoing flow rates in steady regime. For ease, we start with $\beta = \beta_1 = \beta_2$  and generalize the result to $\beta_1\not= \beta_2$ thereafter. We get: 
\begin{equation}
\mathbb{E}(i) = {\mu l \over {\mu \gamma - \mu \beta - \gamma \beta}},
\label{eq:i_mean}
\end{equation}
which, to be a positive and finite quantity, yields the stability criterion in buzz-free regime:
\begin{equation}
{\beta}^{-1} > {\mu}^{-1} + {\gamma}^{-1}. 
\label{eq:stability}
\end{equation}
We now extend these results to the case where the model may exhibit a buzz activity. As $\beta$ alternates between the hidden states $\beta=\beta_1$ and $\beta=\beta_2$, \roy{}{with respective state probabilities} $a_2/(a_1+a_2)$ and $a_1/(a_1+a_2)$, one can simply replace $\beta$ in Eq. (\ref{eq:i_mean}) and (\ref{eq:stability}) with the equivalent average  value: 
\begin{equation}
\bar{\beta} =  {{\beta_{1} a_2} \over {a_1 + a_2}} + {{\beta_{2} a_1} \over {a_1 + a_2}}.
\end{equation}
In order to illustrate the flexibility of our workload model and to validate Eq. (\ref{eq:i_mean}), we generate three synthetic traces  corresponding to the different sets of parameters verifying the stability condition of relation (\ref{eq:stability}) and reported in Table \ref{table1}. Particular realizations of these processes generated over $2^{21}$ points are displayed in  Figure \ref{fig:traces}. }
\begin{table}[h]
\centering
\caption{Parameters value used in the workload model to generate the three  traces plotted in Fig. \ref{fig:traces}. The last two rows correspond to the theoretical mean workload of Eq. (\ref{eq:i_mean}) and to the sample mean value estimated from the traces.}
\begin{tabular}{lccc}
\hline
   & case (a) & case (b) & case (c) \\ \hline
$\beta_1$ & $4.762\times10^{-4}$ & $3.225\times10^{-5}$ & $2.439\times10^{-5}$ \\ 
$\beta_2$ & $0.0032$ & $0.0032$ & $0.0032$ \\ 
$\gamma$ & $0.0111$ & $0.0020$ & $ 0.0011$ \\ 
$\mu$ & $5\times10^{-4}$ & $3.289\times10^{-5}$ & $2.5\times10^{-5}$\\
$l$ & $10^{-4}$ & $10^{-4}$ & $10^{-4}$\\
$a_1$ & $10^{-7}$ & $10^{-7}$ & $10^{-7}$\\
$a_2$ & $0.0667$ & $0.0667$ & $0.0667$ \\[1mm] \hline
$\mathbb{E}(i)$ & $1.92$ & $15.68$ & $44.72$ \\ 
Emp. mean $\langle i\rangle$ & $1.74$ & $16.72$ & $45.23$ \\
\hline
\end{tabular}
\label{table1}
\end{table}
\paulo{}{While the synthetic traces corresponding to  cases (b) and (c) reproduce distinct and easily identifiable buzz regimes, the parameter set of case (a) leads to a workload variation distinct from the typical shape of Figure \ref{fig:viral}. Nonetheless, for all 3 configurations, the empirical means estimated from the $2^{21}$ samples of the traces are in  good agreement with the expected values of \roy{relation}{Eq.} (\ref{eq:i_mean}).
}
\newline
\paulo{It is to be noted that even though we consider exponential distribution in our model for simplicity, any other distribution can be used here and the equations for the mean workload and stability condition (Eq. \ref{eq:i_mean} and \ref{eq:stability}) would remain unchanged. However, our parameter identification procedure is based on exponential assumptions and needs to be adapted accordingly, if other types of distribution were to be used.}
{Finally, let us notice that even though we consider exponentially distributed random variables in our model, any other distributions could be used, which, according to the same balance principle, would lead to a mean workload  and to a stability condition of the same kind as  (\ref{eq:i_mean}) and (\ref{eq:stability}). However, the estimation procedure we derive in the next section strongly relies on the exponential assumption and it would need to be thoroughly reworked to adapt to different hypotheses.
}

\section{Estimation procedure}
\label{sec:estimation}
In this section, we address the identifiability of our model and design a calibration algorithm to fit workload data.
We start constructing empirical estimators for each parameter of the model and we numerically evaluate their performance on synthetic traces. 

\subsection{Parameters estimation}
\label{sec:est_method}

Considering a standard epidemic process $X$ with propagation rate $\theta$, the maximum likelihood estimate $\widehat{\theta}_{\scriptscriptstyle \rm MLE}$ is derived in \cite{BarDynaPro2008}, \cite{AndStoEpi2000} and reads:
\begin{equation}
\widehat{\theta}_{\scriptscriptstyle\rm MLE} =  n \cdot \left( \int_0^T X(t) \,\mathrm{d}t\right)^{-1},
 \label{eq:mle}
\end{equation}
where $n$ is the number of contaminations (i.e. number of increments of $X$) occurring within the time interval $T$.

Very often, maximum likelihood approach yields optimal results (in terms of estimate variance and or bias) but it is not always possible to get a closed-form expression for the estimated parameters. This can either be due to the likelihood function that is impossible to derive analytically, or to missing data that preclude straightforward application of the maximum likelihood principle. Nonetheless, solutions, such as the Expectation-Maximization (EM) or the Monte Carlo Markov Chain (MCMC) algorithms exist, which in some cases can approximate maximum likelihood estimators. 

Returning to our model depicted in Figure \ref{fig:Markov}, each parameter needs to be empirically estimated, assuming that the instantaneous workload time series is the only available observation.\\

\noindent{\it Watching parameter $\gamma$.} As $\gamma$ is the departure rate of users that leave the infected state after they finished watching a video, it can directly be inferred from the number $n$ of decrements of the observable process $I(t)$. Therefore,  the MLE  of \roy{equation}{Eq.} (\ref{eq:mle}) straightforwardly applies and leads to:
\begin{equation}
\widehat{\gamma}_{\scriptscriptstyle\rm MLE} =  n \cdot \left( \int_0^T I(t) \,\mathrm{d}t\right)^{-1}.
 \label{eq:gamma}
\end{equation}
 
\begin{figure}[h]
\centering
\hspace*{-2mm}\begin{tabular}{cc}
\begin{turn}{90}{\hspace*{14mm}$T_{\mu}$} \end{turn} &
\hspace*{-3mm}\includegraphics[width=.55\columnwidth]{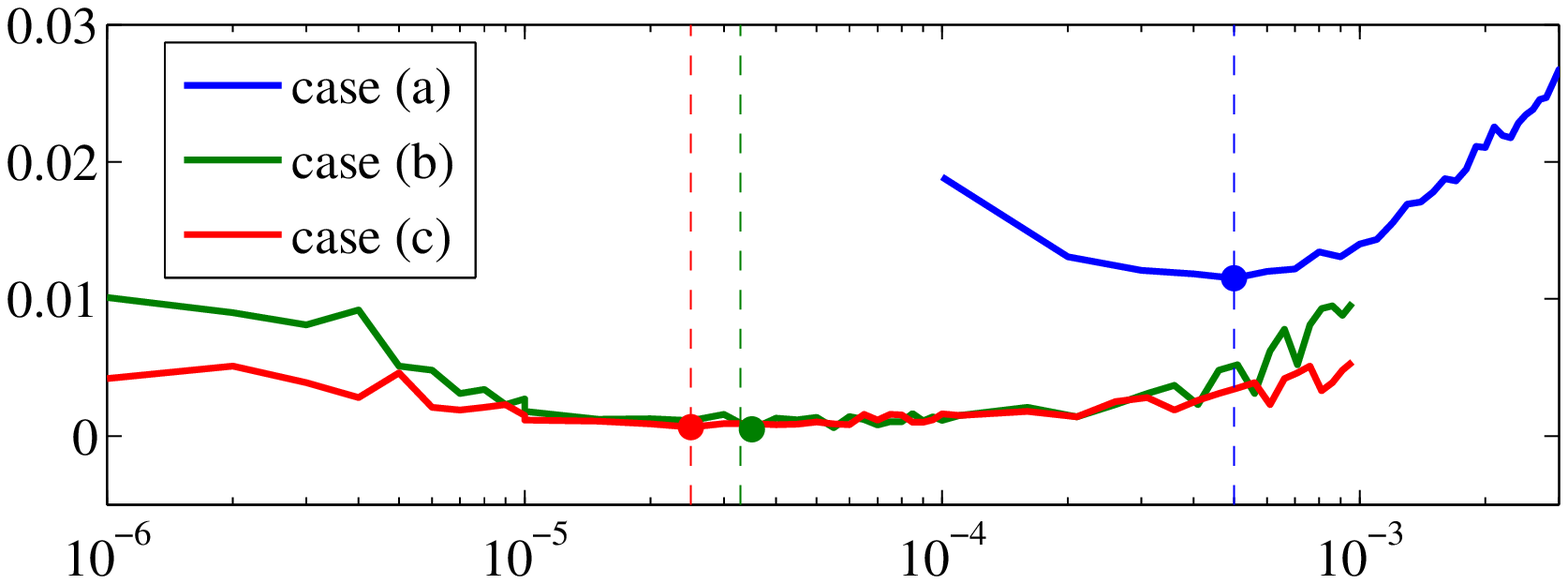} \\[-1mm]
& $\mu$ ({\em logarithmic scale}) \\
\end{tabular}
\caption{\small Evolution of the exponential test statistics (\ref{eq:KS}) applied to the traces of Figure \ref{fig:traces}. Dotted vertical lines locate the actual value of $\mu$ for each case;  dot markers on each curve indicate the estimated value $\widehat{\mu}$ corresponding to the minimum point of the statistical test $T_{\mu}$. }
\label{fig:KS}
\end{figure}

\begin{figure}[h]
\centering
\hspace*{-2mm}\begin{tabular}{cc}
\begin{turn}{90} \hspace*{8mm} $R(t)$, $\widehat{R}(t)$ \end{turn} & 
\hspace*{-3mm}\includegraphics[width=.55\columnwidth]{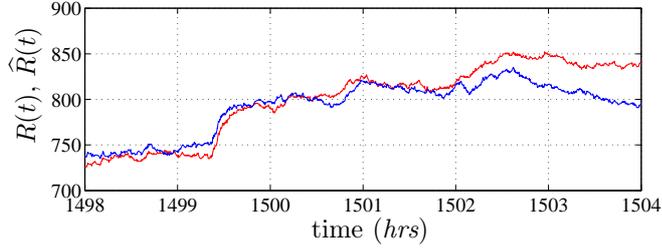} \\[-1mm]
& time ({\em hrs}) \\
\end{tabular}
\caption{\small Evolution of the number of active past viewers. Comparison of the actual (non observable) process $R(t)$ (blue curve) with the estimated process $\widehat{R}(t)$ (red curve) derived from  expression (\ref{eq:estimate-R}).}
\label{fig:R}
\end{figure}

\noindent{\it Memory parameter $\mu$.} This rate at which past viewers leave the recovery compartment and stop propagating the virus (gossip), relates to the decrement density of the non-observed process $R(t)$. It is thus impossible to simply apply the MLE of Eq. (\ref{eq:mle}) unless we first construct a substitute $\widehat{R}(t)$ to the missing data from the observable data set $I(t)$. Let us recall that in our model, all current viewers turn and remain contagious for a mean period of time $\gamma^{-1}+\mu^{-1}$. Then, in first approximation, we can consider that $R(t)$ derives from the finite memory cumulative process: 
\begin{equation}
\widehat{R}(t) = \int_{t-(\gamma^{-1}+\mu^{-1})}^{t} I(u)\, \mathrm{d}u,
\label{eq:estimate-R}
\end{equation}
which itself, depends on the parameter to be estimated $\mu$. 
We propose an estimation procedure based on the inherent exponential property of the model. From the Poisson assumption, the inter-arrival time $\mathbf{w}$ between the consecutive arrivals of two new viewers is an exponentially distributed random variable such that  $\mathbb{E}\,(\mathbf{w}|\,I(t)+R(t)=x) = (\beta\,x+l)^{-1}$. 
\paulo{Our rationale is based on the fact that for a given value of $\mu$, we can evaluate its likelihood with regard to the trace. First, we consider a value of $\mu$ and derive $\widehat{R}(t)$ from Eq. (\ref{eq:estimate-R}). For each value of the sum $I(t)+\widehat{R}(t)$, we obtain a set of inter-arrival times. We normalize them by dividing the samples of each set by their corresponding mean. Then we merge all sets to build a single set of inter-arrival times. Clearly this set contains samples $X_1, X_2 . . . , X_n$ of a random variable $X$, that are exponentially and identically distributed, i.e. they follow the same exponential distribution. We rearrange this set in an ascending order with $X_{(0)}$ as the minimum value followed by $X_{(1)}, . . . X_{(n)}$. We then perform the ``normalized spacings" \cite{Jammalamadaka03} and generate the transformed variable $Y$, such that $Y_{(i)} =(n - i+1)(X_{(i)} - X_{(i - 1)})	, i=1,...,n$. $Y$ is also exponentially and identically distributed, provided that the original variable ($X$) also holds the same property \cite{Seshadri69}. \newline
Following \cite{Jammalamadaka03} we compare the cumulative empirical distribution function (cedf) of the original variable $X$ with that of the transformed one, $Y$. Let $F_n(t)$ and $G_n(t)$ denote the empirical distribution functions	of	$X$ and $Y$ respectively. We then construct test of exponentiality by measuring the distance between these two cedf’s, using the classical Kolmogorov-Smirnov type distances. For each value of $\mu$, we obtain a test statistics ($T_n$) that can be expressed as follows:
\begin{equation}
T_{n} = \sqrt{\frac{1}{n}}	 \sup_{0 \leq t \leq \infty} |F_n(t) - G_n(t)|
\label{eq:estimate-mu}
\end{equation}
$T_n$ somehow expresses the likelihood of $\mu$ as being the right value that generates the workload trace.
For $\mu \in [\epsilon, \gamma]$ (Typically we choose $\epsilon$ in the order of $10^{-7}$ with steps of $2 \times 10^{-7}$) we roughly reconstruct $\widehat{R}(t)$ and compute $T_n$ following Eq. (\ref{eq:estimate-mu}). We choose our estimated value of $\mu$, i.e. $\widehat{\mu}$ as the one with the lowest value of $T_n$. In Figure \ref{fig:KS}  we show how the value of $T_n$ varies with $\mu$ for the three cases. It clearly indicates a minimum value of $T_{n}$ which is around the actual value of $\mu$.}
{
Its means that, for $x$ fixed, the normalized random variable $\widetilde{\mathbf{w}} = \mathbf{w}/\mathbb{E}(\mathbf{w}|x)$ is exponentially distributed with unitary parameter and becomes independent of $x$. Ideally then, for each value of $R(t)+I(t)=x$,  all the sub-series  ${\rm w}_x =\{w_n~:~R(t_n)+I(t_n)=x\}$, after normalization by their own empirical mean, yield independent and identically distributed realizations  of a unitary exponential random variable. In practice though, as $R(t)$ is not observable, only if  $\widehat{R}(t)$ is accurately estimated, should this unitary exponential i.i.d. assumption hold true. From there, we propose the following algorithm: for different values of $\mu$ spanning a {\it reasonable} interval, we use $\widehat{R}_{\mu}(t)$ estimated from \roy{}{Eq.} (\ref{eq:estimate-R}) to build the normalized series $\widetilde{\rm w}_{\mu}$. A statistical test applied to each $\widetilde{\rm w}_{\mu}$ allows for assessing the exponential i.i.d. hypothesis and then to select the value of $\mu$ that yield the best score. \\
More concretely, we apply to $\widetilde{\rm w}_{\mu}=\left(\widetilde{w}_n\right)_{n=1,\ldots,N }$ the statistical exponentially test derived in \cite{Jammalamadaka03}: Form the {\it normalized spacings} ${\rm v}_{\mu} = \left(v_{(n)} =(N - n+1)(\widetilde{w}_{(n)} - \widetilde{w}_{(n - 1)})\right)_{ n=1,...,N}$ where $\left(\widetilde{w}_{(n)}\right)_{n=1,\ldots, N}$ stands for $\widetilde{\rm w}_{\mu}$ rearranged in ascending order. Let $F$ and $G$ denote the cumulative distribution functions of $\widetilde{\rm w}_{\mu}$ and  ${\rm v}_{\mu}$ respectively, and compute the  classical Kolmogorov-Smirnov distance:
\begin{equation}
T_{\mu} = \sqrt{\frac{1}{N}}	 \sup_{1 \leq k \leq N} |F(k) - G(k)|.
\label{eq:KS}
\end{equation}
As $F$ and $G$ are identical for an exponentially i.i.d. random series, we then expect $T_{\mu}$ to reach its minimum  for the value of $\mu$ that gives the best estimate $\widehat{R}_{\mu}(t)$ of $R(t)$:
\begin{equation}
\left\{\begin{array}{l}
\widehat{\mu} = {\rm argmin}_{\mu} T_{\mu} \\[1mm]
\widehat{R} = \widehat{R}_{\widehat{\mu}}.
\end{array}\right.
\label{eq:mu-estimate}
\end{equation}
Plots of Figure \ref{fig:KS} show the evolution of the Kolmogorov-Smirnov distance corresponding to the traces displayed in Figure \ref{fig:traces}. In the 3 cases, $T_{\mu}$ clearly attains its minimum bound for $\widehat{\mu}$ close to to the actual value. The corresponding estimated processes $\widehat{R}(t)$ derived  from \roy{}{Eq.} (\ref{eq:mu-estimate}) match fairly well the real evolution of the ({\bf R}) class in our model (see Figure \ref{fig:R}).\\

 }

\begin{figure}[h]
\centering
\hspace*{-3mm}\begin{tabular}{cc}
\begin{turn}{90}\hspace*{25mm} $(\Omega(x))^{-1}$ \end{turn}&
\hspace*{-2mm}\includegraphics[width=0.45\columnwidth]{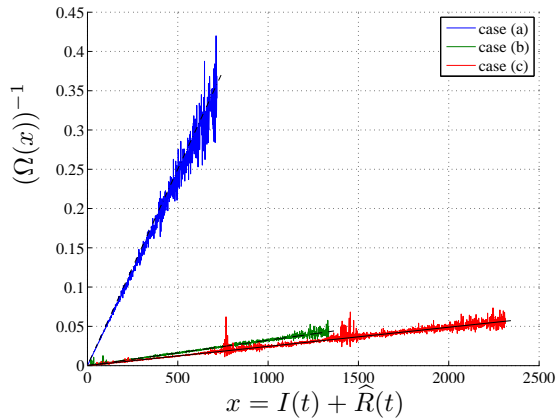} \\[-1mm]
& $x = I(t)+\widehat{R}(t) $ \\
\end{tabular}
\caption{\small Weighted linear regression of $\Omega^{-1}$ vs $x$ corresponding to the three traces of Figure \ref{fig:traces}. Superimposed are the  linear trends fitted on the respective data.}
\label{fig:regression}
\end{figure}

\noindent{\it Propagation parameters $\beta$ and $l$.} 
According to our model, the arrival rate $\lambda(t)$ of new viewers is given by \roy{equation}{Eq.} (\ref{eq:infinity}). It linearly depends on the current number of active and past viewers. So, from the observation $I(t)$ and the reconstructed process $\widehat{R}(t)$ of Eq. (\ref{eq:mu-estimate}), we could formally apply the maximum likelihood \roy{equation}{Eq.} (\ref{eq:mle}) to estimate $\beta$. In practice however, we have to bear in mind that: {\it (i)}  the arrival process of rate $\lambda(t)$  comprises a spontaneous viewers ingress that is governed by parameter $l$ and which is independent of the current state of the system; {\it (ii)} depending on the current hidden state of the model (buzz-free  {\it versus} buzz state), it is alternately $\beta=\beta_1$ and $\beta=\beta_2$ that fix the propagation rate in Eq. (\ref{eq:infinity}). 
We designed an estimation procedure based on a weighted linear regression, that simultaneously addresses these two issues.  We decompose our rationale in two steps: First, let us consider the buzz-free state only and $\beta=\beta_1$. As discussed in the estimation of $\mu$ the inter-arrival time $\mathbf{w}$ between the consecutive arrivals of two new viewers is an exponentially distributed random variable such that  $\mathbb{E}\,(\mathbf{w}|\,I(t)+R(t)=x) = (\beta\,x+l)^{-1}$. Concretely then, for different values of the sum $I(t)+\widehat{R}(t)$, we calculate the  conditional empirical mean:
\begin{equation}
{\Omega}(x) = \frac{1}{|\mathcal{I}(x)|}\,\sum_{t_n\in\mathcal{I}(x)} w_{n}~:~\mathcal{I}(x)=\{t_n:I(t_n)+\widehat{R}(t_n)=x\}.
\label{eq:linear-regression}
\end{equation}
The linear regression of $(\Omega(x))^{-1}$ against $x$ yields at one go, both parameters estimation $\widehat{\beta}$ (slope) and $\widehat{l}$ (intercept).\\
Let us now return to the general form of our model with alternation of buzz and  buzz-free periods. In the buzz-free case, $\beta=\beta_1$ corresponds to a normal workload activity, meaning that the sum $I(t)+\widehat{R}(t)$ takes on rather moderate values. Conversely, when the system undergoes a \roy{flash crowd}{buzz}, $\beta=\beta_2$ and the population $I(t)+\widehat{R}(t)$ suddenly increases to reach significantly larger values. Yet, in both cases, the quantity $\Omega^{-1}$ defined in Eq. (\ref{eq:linear-regression}) remains linear with $x$ but with two different regimes (slopes) depending on the amplitude of $I(t)+\widehat{R}(t)=x$. As a result, 
it is possible to reduce the bias that  $\beta_2$ causes on the estimation of $\beta_1$, using a weighted linear regression of $\Omega^{-1}$ {\it vs} $x$ where the weights $p(x)$ are proportional to the cardinal of the 
indicator sets $\mathcal{I}(x)$. Indeed, $|\mathcal{I}(x)|$ should be smaller for larger values of $x$ because buzz episodes are expected to be less frequent than nominal activity periods. Figure \ref{fig:regression} confirms the claim:  the plots $(x,\Omega^{-1})$ show a manifest linear trend with higher variability at $x$'s large, meaning a fewer terms entered the sum of \roy{expression}{Eq.} (\ref{eq:linear-regression}).

Formally, we can apply the exact same procedure to estimate $\beta_2$, but considering opposite weights to favor 
the large values of $x$'s. However, due to the large fluctuations of $(\Omega(x))^{-1}$ in the corresponding region, the  slope $\widehat{\beta_2}$  is subject to a very poor estimation variance. Instead, we propose to apply the ML estimator \roy{}{described in Eq.} (\ref{eq:mle}) on the restriction of $I(t)$ to the buzz  periods only. Strictly speaking, we should consider $\widehat{R}(t)$ as well, but since a buzz event normally occurs on very small interval of time, we assume that  $\widehat{R}(t)$ (resp. $R(t)$) remains constant in the meanwhile (flash crowd viewers will enter in  R compartment only after the visualization time). In practice, to automatically identify the buzz periods, we threshold $I(t)$ and consider only the persistent increasing parts that remain above the threshold.\\

\noindent{\it Transition rates $a_1$ and $a_2$.} As we already said, at time $t$, the inter-arrival time $\mathbf{w}$ separating to new incomers is a random variable drawn from an exponential law of parameter $\lambda = \beta(i+r)+l$, where $I(t)+R(t)=i+r$ and $\beta$ is either equal to $\beta_1$ or to $\beta_2$. We denote $f_1(\mathbf{w})$ and $f_2(\mathbf{w})$ the corresponding densities built upon the reconstructed process $\widehat{R}(t)$ and  the estimated parameters $(\widehat{\beta_1}, \widehat{l})$ and $(\widehat{\beta_2}, \widehat{l})$ respectively.  For a given inter-arrival time $\mathbf{w}=w_n$ observed at time $t_n$, we form the likelihood ratio $ f_2(w_n)/f_1(w_n) $ to determine whether the system is in buzz or in buzz-free state. Moreover, in order to avoid non-significant state transitions we resort to a restoration method inspired by the Viterbi algorithm \cite{Kleinberg2002}.
Once we have identified the hidden states of the process, we estimate the transitions rates $\widehat{a_1}$ and $\widehat{a_2}$ from the average times spent in each state.


\subsection{Numerical Validation}
\label{sec:est_results}

\begin{figure}[h]
\centering
\hspace*{-5mm}\begin{tabular}{ccc}
\hspace*{5mm}case (a) & \hspace*{5mm}case (b) & \hspace*{5mm}case (c)\\[-1mm]
\includegraphics[width=0.31\columnwidth]{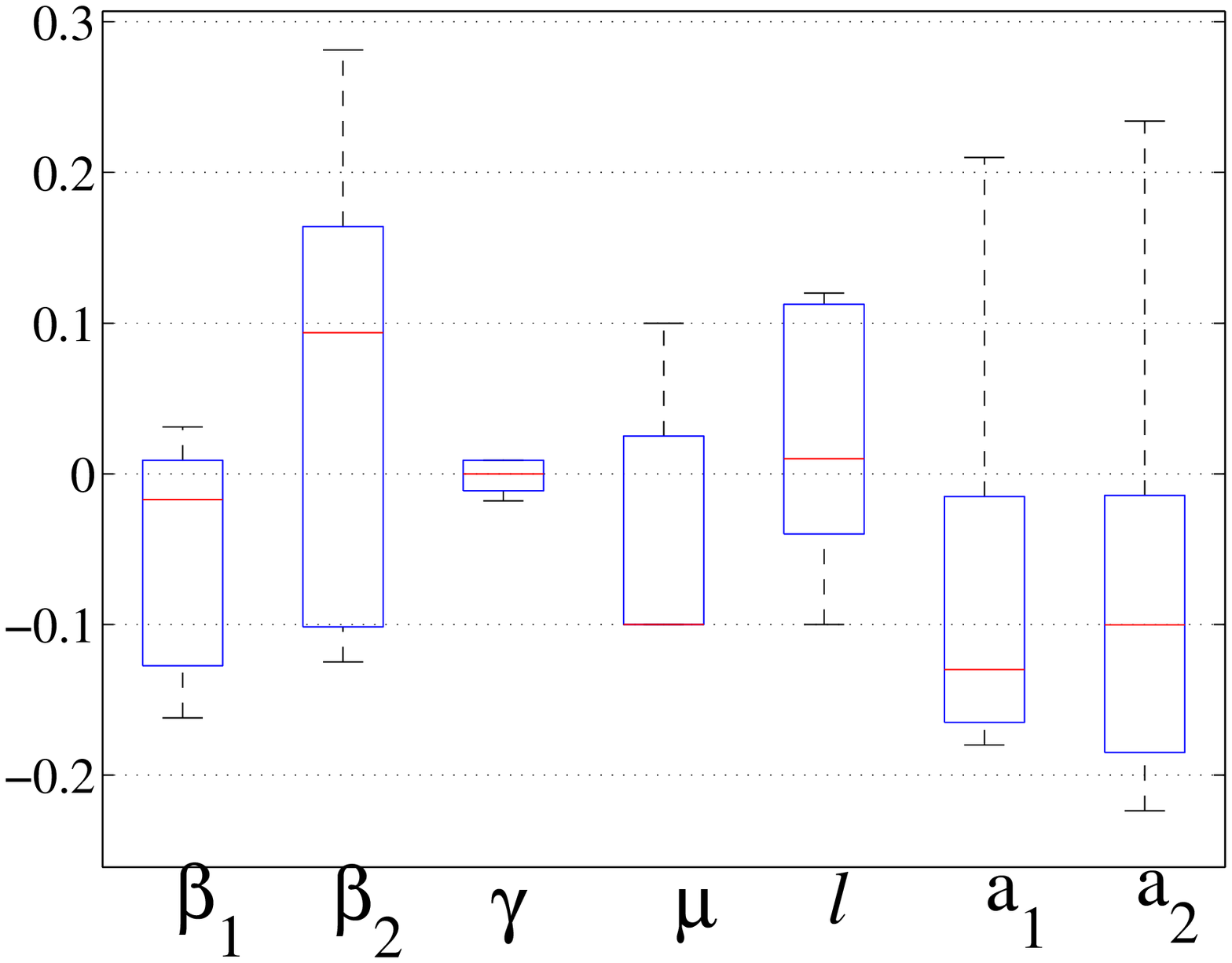} &
\includegraphics[width=0.31\columnwidth]{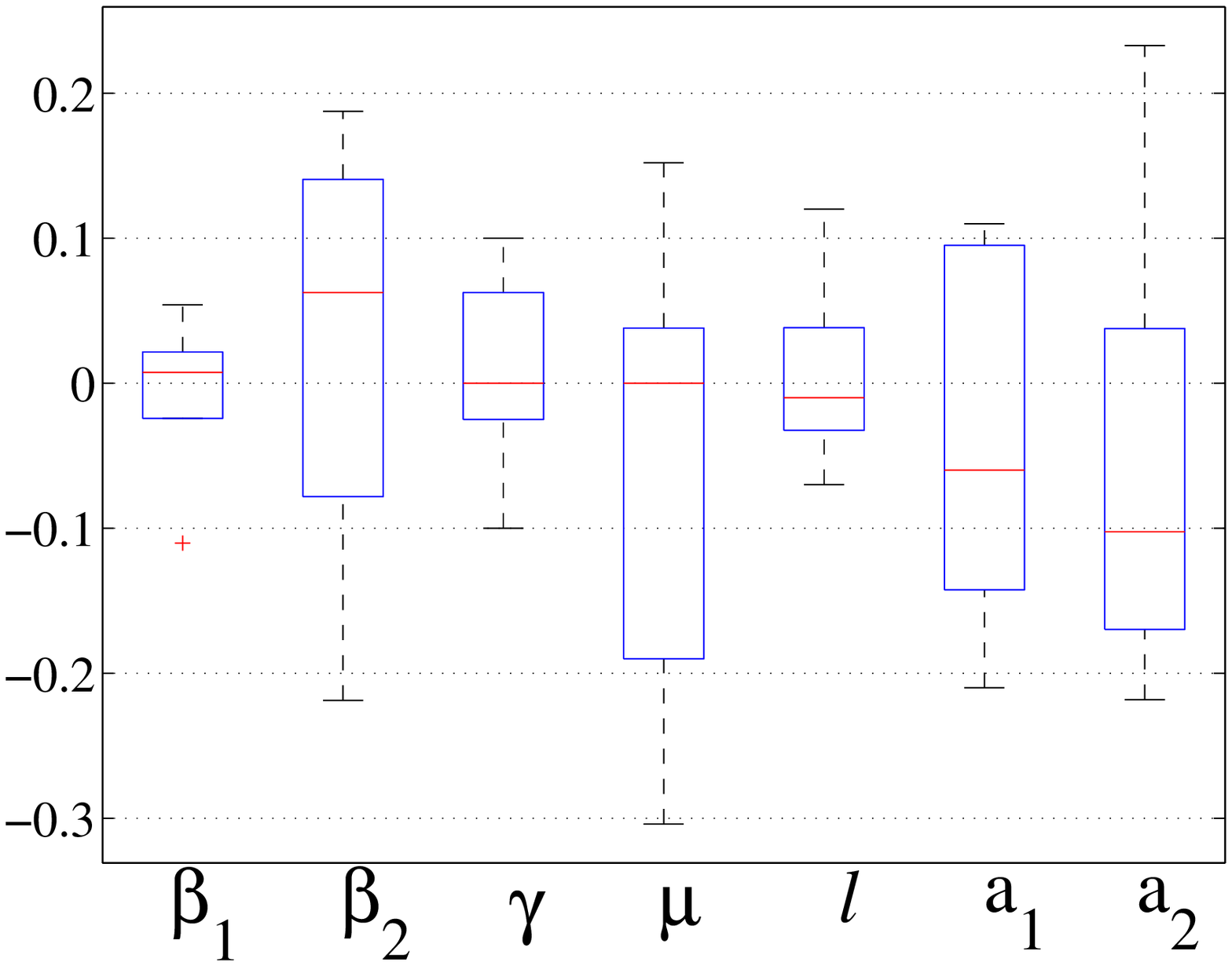} &
\hspace*{3mm}\includegraphics[width=0.29\columnwidth]{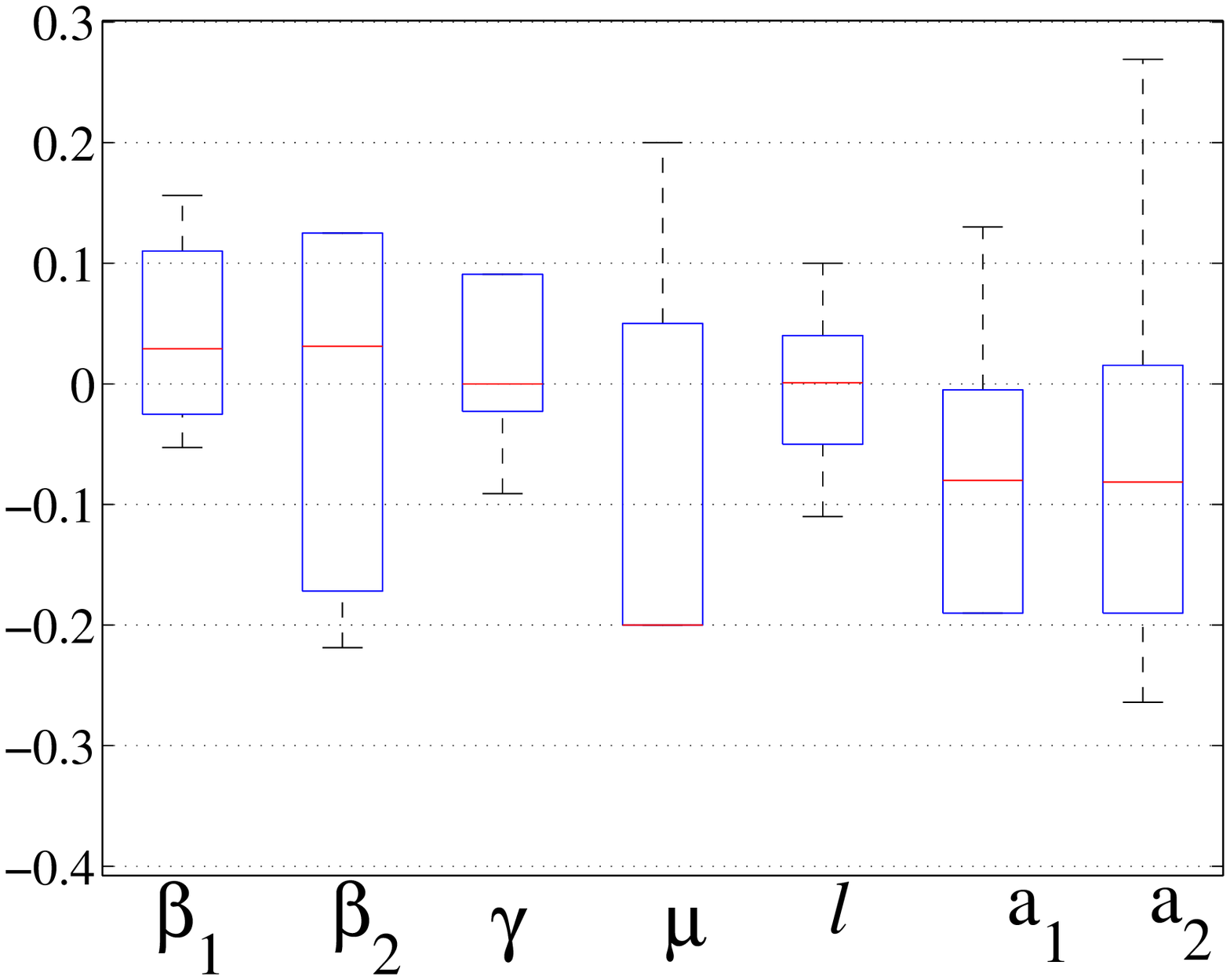} 
\end{tabular}
\caption{\small Box-and-Whisker plots for relative estimation errors of the model parameters for the three different sets of prescribed parameters reported in Table \ref{table1}. For each case (a)-(c), statistics are computed over 10 independent realizations of time series of length  $2^{21}$ points.}
\label{fig:box-plot}
\end{figure}

\begin{figure}[h]
\centering
\hspace*{-5mm}
\begin{tabular}{cc}
\begin{turn}{90} \hspace*{20mm} $\log_2 {\rm MSE}(\widehat{\theta})$ \end{turn}&
\includegraphics[width=0.45\columnwidth]{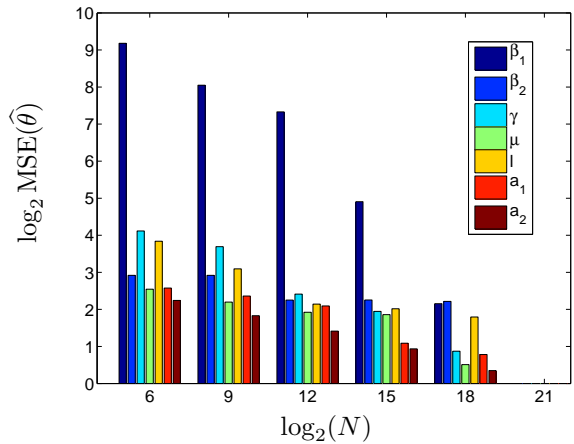} \\
& $\log_2(N)$ \\
\end{tabular}
\caption{\small Evolution of the Mean Square Error {\em versus} the data length $N$ in a {\em log-log} plot. For the sake of conciseness, we  only show here the results corresponding to the case (b) of Table \ref{table1}.
}
\label{fig:GB}
\end{figure}

\paulo{We validate our estimation procedure against the synthetic traces with three different workloads as shown in Figure \ref{fig:traces}. Each trace contains $2^{21}$ events and we conducted $10$ independent realizations to produce $10$ independent traces for each case. Then we apply our estimation procedure on each of trace. For each parameter we obtain the so called ``descriptive statistics", i.e. the smallest observation (sample minimum), lower quartile, median, upper quartile, and largest observation (sample maximum) from the box-and-whisker plot. \newline
From Figure \ref{fig:box-plot} we can infer that estimation of $\gamma$ is almost unbiased with an interquartile range (IQR) less than $10\%$ for all three cases. $\beta_1$ is estimated with little bias but $\beta_2$ is harder to estimate and the IQR ranges around $15\%$. As we discussed in the previous section, estimation of $\mu$ is not trivial. We find that this estimation is negatively biased for the three cases with IQR around $20\%$. Estimation of $a_1$ and $a_2$ are also negatively biased with IQR less than $20\%$.}
{
To evaluate the statistical performance of our estimation procedure, we resort to numerical experiments to empirically get the first and the second order moments of each parameter estimator. Owing to  the versatility of our model, we must ensure that the proposed calibration algorithm performs well for a variety of workload dynamics. To this end, we systematically reproduce the experiments considering the 3 sets of parameters reported in Table \ref{table1}. For each configuration, we generate 10 independent  realizations of processes similar to the ones depicted in Figure \ref{fig:traces}, and use these to derive  descriptive statistics. 

The box-and-whisker plots of Figure \ref{fig:box-plot} indicate for each estimated parameter (centered and normalized by the corresponding actual value) the sample median (red line), the inter-quartile range (blue box height) along with  the extreme samples (whiskers) obtained from time series of length $2^{21}$ points each. As expected (owing to the maximum likelihood procedure), estimation of $\gamma$ shows to be the most accurate, both in terms of bias and variance. But more surprisingly though, although the estimation $\widehat{\beta_1}$ derives from a heuristic procedure that  itself depends on the raw approximation $\widehat{R}(t)$ of Eq. (\ref{eq:estimate-R}), the resulting performance is remarkably good: bias is always negligible (less than 5\% in the worst case (c))  and the variance always confines to 10\% of the actual value of $\beta_1$. Notice also that the estimation of $\beta_1$ goes from a slight underestimation in case (a) to a slight overestimation in case (c), as the buzz effect,  i.e. the value of $\beta_2$, grows from traces (a) to (b). Compared to $\widehat{\beta_1}$, the estimation of ${\beta_2}$ behaves more poorly and proves to be the most difficult parameter to estimate. But we have to keep in mind that this latter is only based on buzz periods which represent only a small fraction of the entire time series. Regarding the parameter $\mu$, its estimation remains within a 20\% inter-quartile range but cases (a) and (c) show a systematic bias (median hits the lower quartile bound). Let us then recall that the procedure\roy{}{, described by Eq.} (\ref{eq:mu-estimate}) to determine $\widehat{\mu}$ selects within some discretized interval, the value of $\mu$ that yields the best $T_{\mu}$ score. It is then very likely that the true value does not coincide with any sampled point of the interval and therefore, the procedure picks the closest one that systematically lies beneath or above. Finally,  estimation of the transition parameters $a_1$ and $a_2$ between the two hidden states relies on all other parameters estimation, cumulating so all relative inaccuracies. Nonetheless and despite a systematic underestimating trend, precision remains within a very acceptable confidence interval. 
}
\paulo{
We also check how the accuracy of our estimation procedure varies with the number of samples for one case (case (b)). We consider $\beta_1$, $\gamma$ and $\mu$ to show the convergence. Figure \ref{fig:GV} shows the relative error of estimation against the number of samples. As a general comment we infer that relative error in the estimation procedure of $\beta_1$, $\gamma$ and $\mu$ decreases with the increase in trace length, i.e. the number of points in the samples. In our estimation procedure the rates of convergence are $0.64, 0.82$ and $0.067$ for $\beta_1, \gamma$ and $\mu$ respectively. We observe that $\beta_1$ and $\gamma$ converge rapidly with increased trace length. However, for $\mu$ the estimation error varies very little over the number of points.
}
{
Convergence rate of the empirical estimators is another important feature that binds the estimate precision to the amount of available data. Using the same data set, the bar plots of Figure \ref{fig:GB} depicts the evolution of the mean square error ${\rm MSE}(\widehat{\theta})=\mathbb{E}\{(\widehat{\theta} - \theta)^2\}$ -- where generic $\theta$ stands for any parameter of the model -- with the length $N$ of the observable time series $I(t)$. As our purpose  is to stress the rate of convergence of these quantities towards zero, to ease the comparison, we normalize the MSE of each parameter by its particular value at maximum data length (i.e $2^{21}$ points here). Then, the estimator rate of convergence $\alpha_{\theta}$ corresponds to the decaying slope of the MSE with respect to $N$ in a {\em log-log} plot, i.e. ${\rm MSE}(\widehat{\theta}) \sim O(N^{-\alpha_{\theta}})$. For the different parameters of our model we obtain convergence rates that lie between $\alpha_{\beta_1}=0.9$ and $\alpha_{a_2}=0.2$, leading each time to sub-optimal convergence ($\alpha_{\theta}  < 1$). It is worth noticing that, despite its relatively {\em ad hoc} construction, the estimator of  $\beta_1$ has an almost optimal convergence rate, which proves the rationality of our approach. 
}

 \newpage

\section{Validation of the estimation procedure against a real workload trace}
\label{sec:validation}
\begin{figure}
\centering
\hspace*{-5mm}\begin{tabular}{cc}
\begin{turn}{90}\hspace*{10mm} Number of current viewers \end{turn} &
\hspace*{-2mm} \includegraphics[width=0.5\columnwidth]{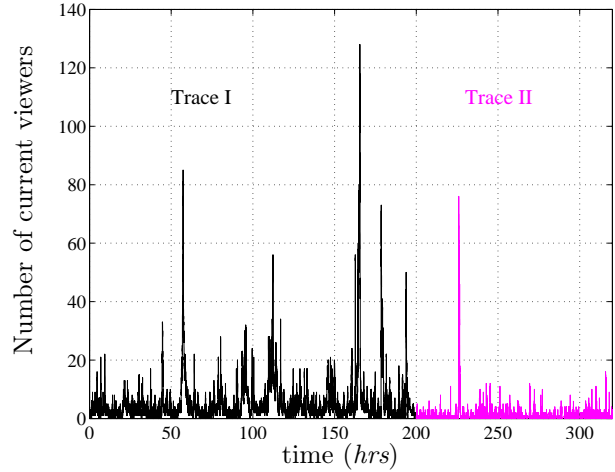} \\[-1mm]
& time ({\em hrs}) \\
\end{tabular}
\caption{\small Real workload time series corresponding to a VoD server demand from \cite{website:grnet}. Initial trace was scaled up by a factor of $10$ to increase the mean workload. Trace is chopped into two separate processes (Trace I and II) corresponding to different activity levels.}
\label{fig:real_trace}
\end{figure}

We now apply the calibration procedure detailed in the previous section, to fit our model on real data and to assess the data-model adequacy in the context of highly volatile workloads. As a paradigm for variable demand applications, we use a VoD trace, released by the Greek Research and Technology Network (GRNET) VoD servers \cite{website:grnet}. Since the trace shows modest activity with a workload level that is not sufficient to stress-test our system, we scale up the data, shrinking all inter-arrival times by a factor of $10$. The resulting workload time series displayed in Figure \ref{fig:real_trace}, clearly shows two distinct periods of steady activity before and after the time index $t=200$. We consider the sub-series on both sides of this cutoff time, as two individual workload processes, referred to as trace I and trace II respectively, and we calibrate our model on each of them separately. 

\begin{figure}[h]
\centering
\begin{tabular}{cc}
Steady-state distribution & Autocorrelation function \\
\includegraphics[width=.4\columnwidth]{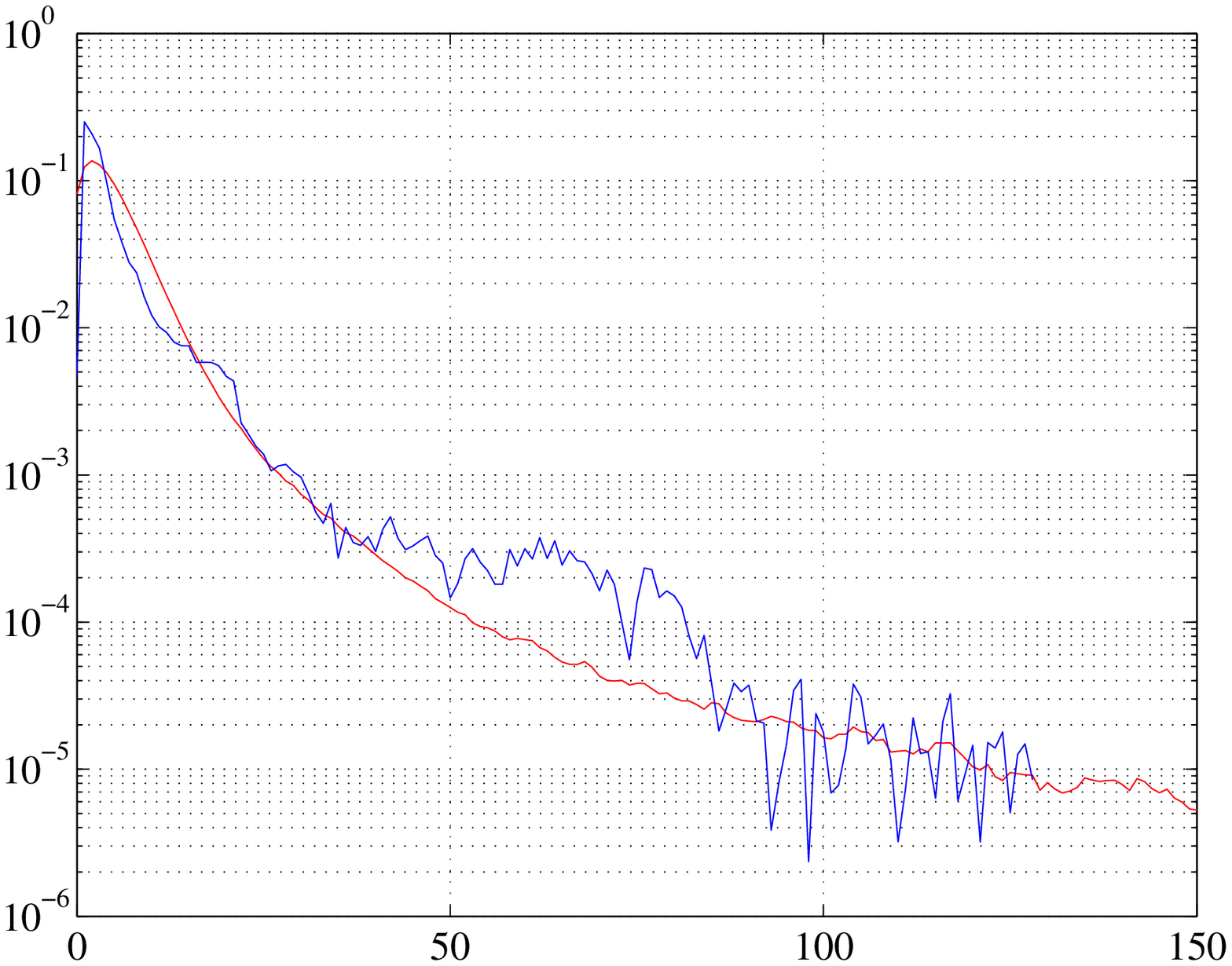} &
\includegraphics[width=.4\columnwidth]{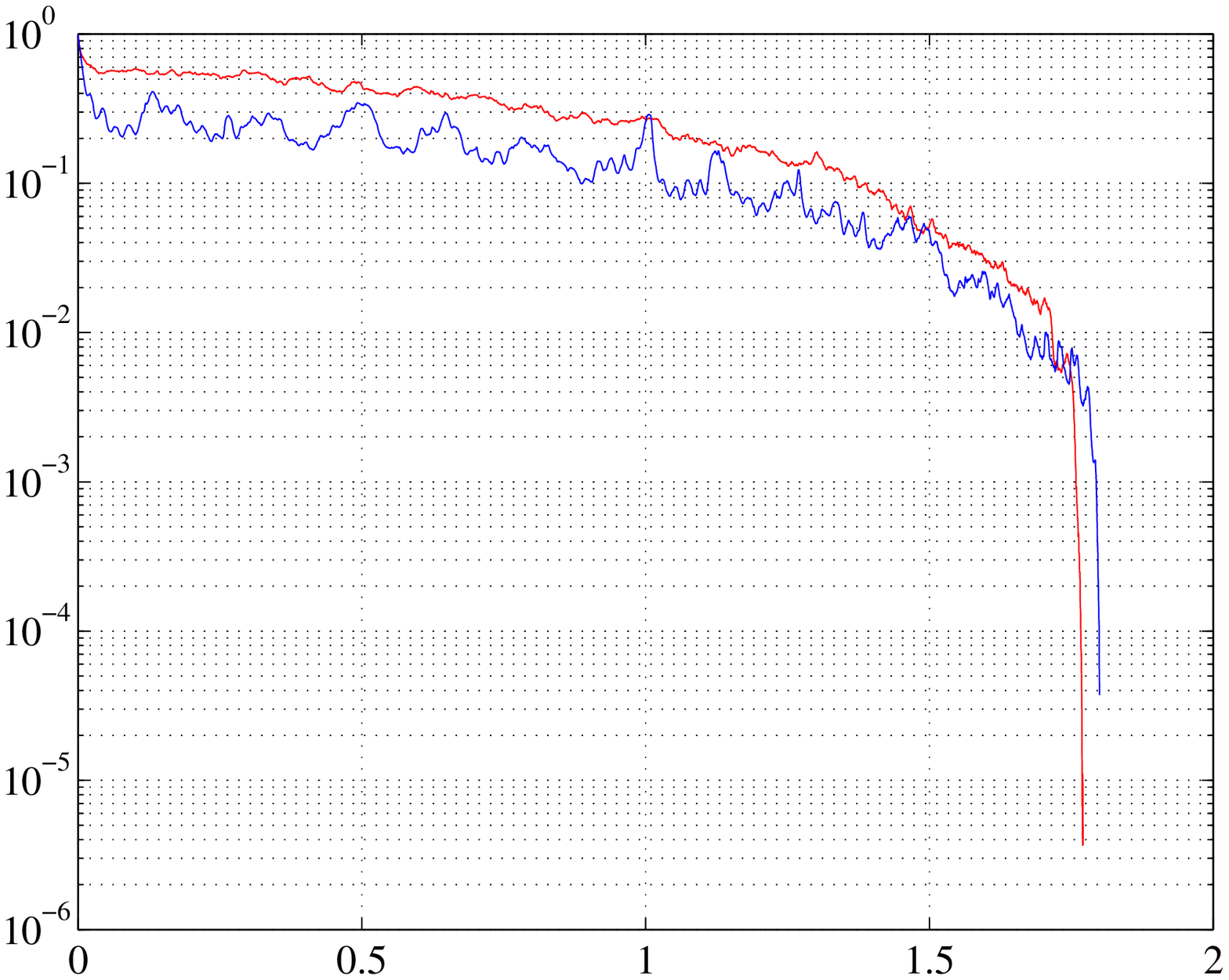} \\
\includegraphics[width=.4\columnwidth]{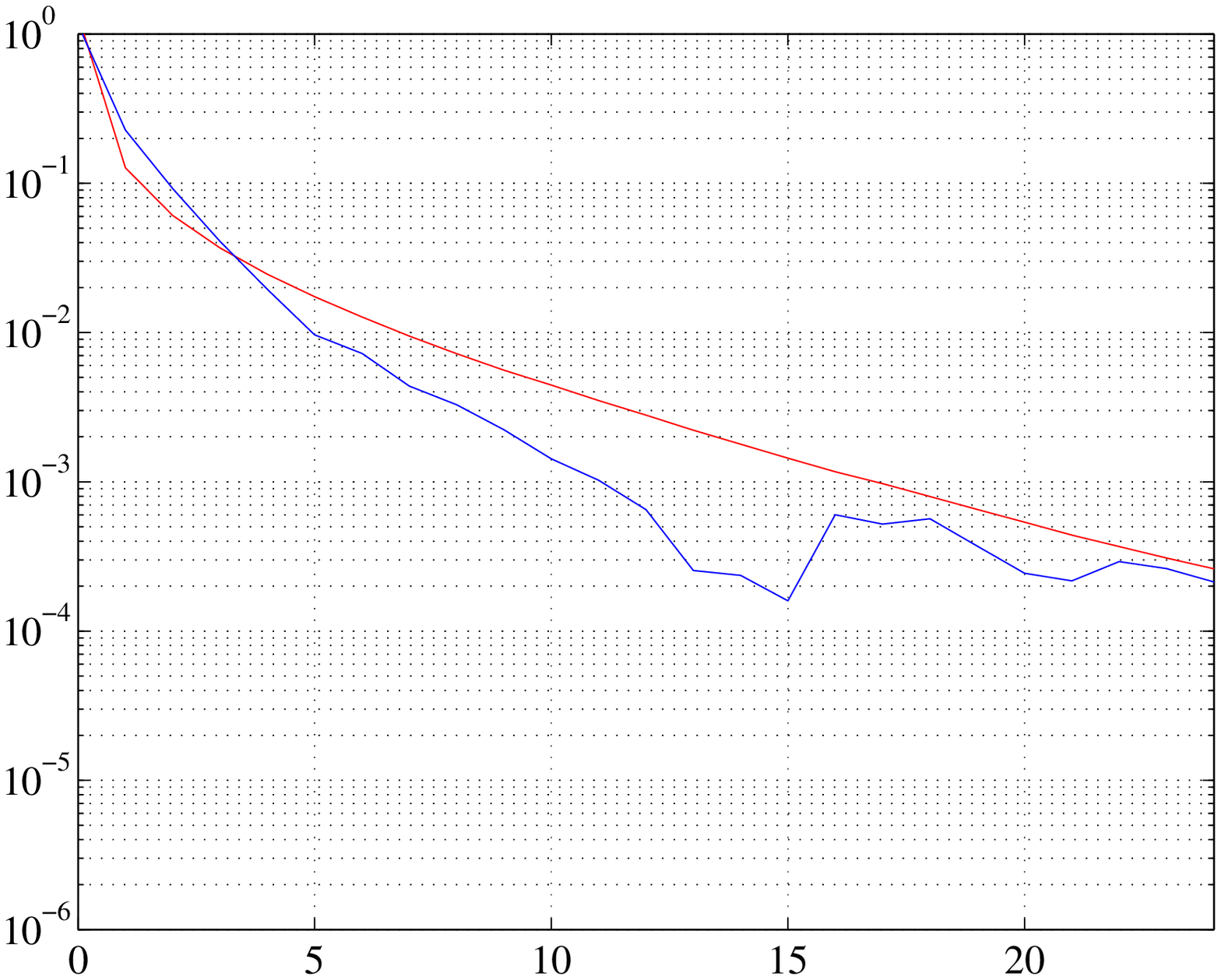} &
\includegraphics[width=.4\columnwidth]{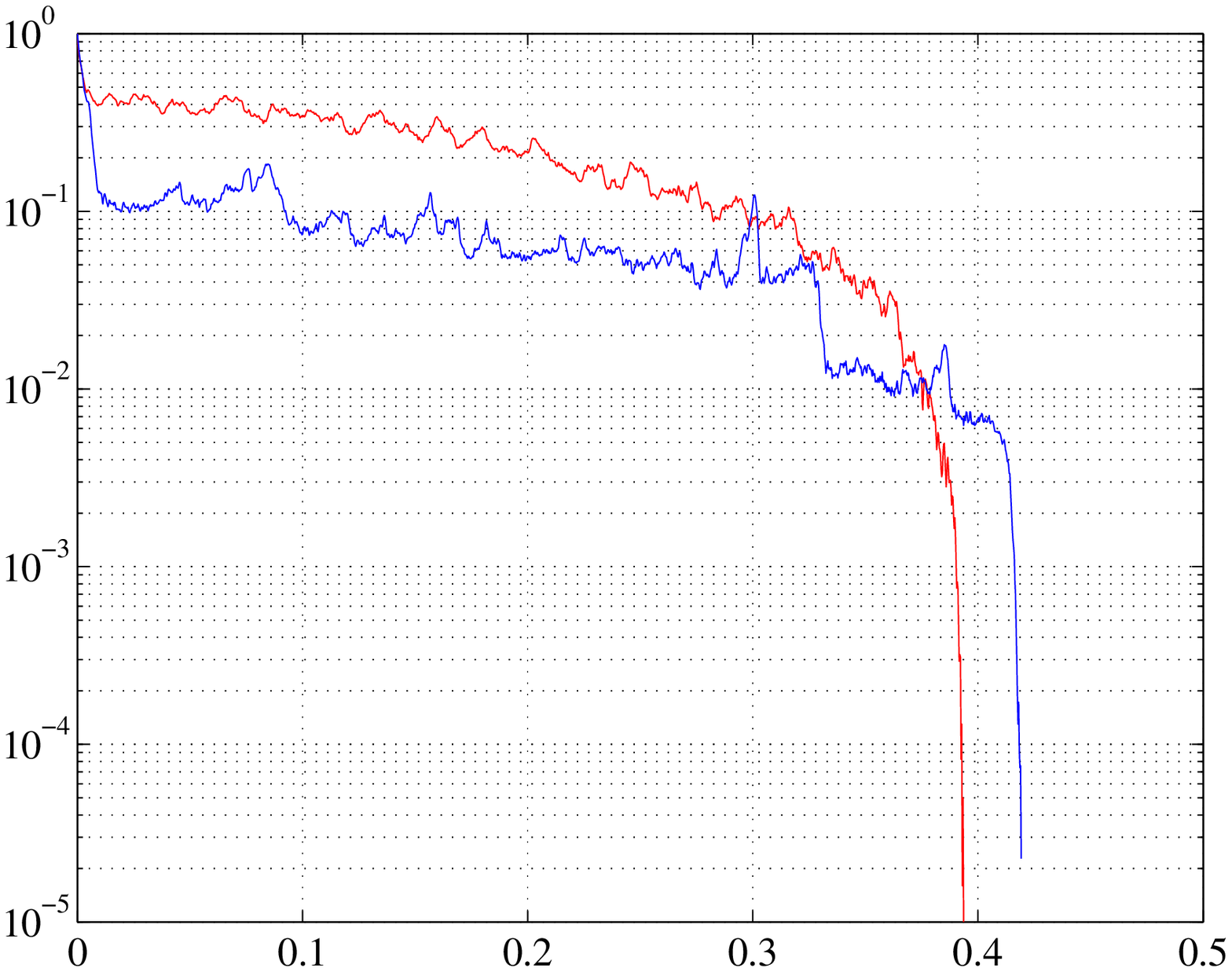}\\
Number of current viewers & time lag $\tau$ ({\em hrs})\\
\end{tabular}
\caption{\small Comparison of the empirical steady-state distribution and of the autocorrelation function of the real (blue curves) and the fitted (red curves) traces. Top two plots correspond to trace~I and bottom plots correspond two trace~II.}
\label{fig:SS&AC}
\end{figure}

Results of the parameters estimation are reported in Table \ref{table2}, and we verified that in both cases the stability condition of Eq. (\ref{eq:stability}) was satisfied. In the same vein, we also compared the empirical mean workload of each trace with its corresponding theoretical value given by the formula (\ref{eq:i_mean}). We obtain for trace I a relative difference of 12\% ($\mathbb{E}(i) = 5.59$ compared to $\langle i \rangle = 4.99$), and of 12.5\% for trace II ($\mathbb{E}(i) = 0.621$ compared to $\langle i \rangle = 0.71$). Naturally, the correspondence here is not as striking as it is with the synthetic traces of Section \ref{sec:vod}. But we must bear in mind that first, {\em ab initio} nothing guarantees that  the underlying system matches our model dynamics and, second, traces I and II can possibly encompass short scale non-stationary periods (e.g. day {\em versus} night activity) which are not accounted for in our model. Notwithstanding this, the match we observe is quite satisfactory and we now focus on higher order statistics to further characterize the data model adequacy. As we do not have a closed-form for the steady state distribution of our Markov process model, nor we have an analytic expression for its autocorrelation function, we use the two sets of  estimated parameters of Table \ref{table2} to synthesize two time series that we compare to the real workload traces I and II. We refer to those synthetic traces as to the fitted traces I and II. The plots in Figure \ref{fig:SS&AC} show the empirical steady state densities and the sample autocorrelation functions of both the real and the fitted traces. The superimposition of the different curves is a clear evidence of our model ability at catching the statistical distribution of the number of current viewers along time. But also, and perhaps more importantly, it demonstrates that the dynamical mechanism underlying our constructive model is able to perfectly reproduce the temporal structure of the real traces, by imposing the correct statistical dependencies between distant samples $I(t)$ and $I(t+\tau)$ of the process.

In addition to serve as a mean to evaluate the goodness-of-fit of our model, the estimated parameters bring on their own, a valuable insight about the system itself. For instance let us compare the propagation rates $\widehat{\beta_1}$ and $\widehat{l}$ estimated from traces I and II, successively. In the first case, $\widehat{\beta_1} < l$, meaning that arrival of new viewers is dominated by spontaneous incomers and is not so much due to information propagation through gossip. Conversely, $\widehat{\beta_1} > \widehat{l}$ for the second workload regime, indicating that the spontaneous attraction of the server has severely dropped whereas the peer-to-peer diffusion component significantly increased but not sufficiently to sustain the mean workload activity. At the same time, the index $\widehat{\mu}$ tripled, meaning that the mean memory period for propagation shrank by a factor of 3. This parameter could then be used as an indicator of the content interest delivered by the server, and of its lifetime in users mind.

\begin{table}[h]
\caption{Estimated Parameters from traces I and II separately.}
\begin{tabular*}{0.83\textwidth}{@{\extracolsep{\fill}} c c  c  c  c  c  c  c  c  c }
  \hline\\[-3mm]
&  $\widehat{\beta_1}$ & $\widehat{\beta_2}$ & $\widehat{\gamma}$ & $\widehat{\mu}$ & $\widehat{l}$ & $\widehat{a_1}$ & $\widehat{a_2}$\\[0.5mm]
  \hline
   I &
  0.0013  & 0.0084  & 0.0039  &  0.0028 & 0.0032 &  $3.13.10^{-4}$ & 0.022\\
  II &
  0.0049  & 0.0183  & 0.0118  &  0.0095 & 0.0005 &  $1.32.10^{-5}$ & 0.041\\ \hline
\end{tabular*} 
\label{table2}
\end{table}
 \newpage
\section{Large Deviation Principle and its interpretation}
\label{sec:ldp}
Consider a continuous-time Markov process $(X_t)_{t\ge0}$, taking values in a finite state space $S$, of rate matrix $A = (A_{ij})_{i\in S, j\in S}$.  In our case $X$ is a vectorial process $X(t) = \left( N_I(t), N_R(t)\right), \forall t\ge0$, and $S = \{ 0, \cdots, I_{\textrm{max}} \} \times \{ 0, \cdots, R_{\textrm{max}} \}$. If the rate matrix $A$ is irreducible, then the process $X$ admits a unique steady-state distribution $\pi$ satisfying $\pi A = 0$. Moreover, by Birkhoff ergodic theorem, it is known that for any mapping $\Phi: S \to \R$, the sample mean of $\Phi(X)$ at scale $\tau$, i.e. $1/\tau\cdot\int_{0}^{\tau} \Phi(X_s) \ud s$  converges almost-surely towards the mean of $\Phi(X)$ under the steady-state distribution, as $\tau$ tends to infinity. The function $\Phi$ is often called the \emph{observable}. In our case, as we are interested in the variations of the current number of users $N_I(t)$, $\Phi$ will simply be the function that selects the first component: $\Phi(N_I(t), N_R(t)) = N_I(t)$. 
The large deviations principle (LDP), which holds for irreducible Markov processes on a finite state space \cite{Varadhan08a}, gives a efficient way to estimate the probability for the sample mean calculated over a large period of time $\tau$ to be around a value $\alpha\in\R$ that deviates from the almost-sure mean:
\begin{equation}
\hspace*{-6mm} \lim_{\epsilon\to0} \lim_{\tau\to\infty} \frac{1}{\tau} \log \PP \left\{  \int_{0}^{\tau} \Phi(X_s) \ud s \in [\alpha - \epsilon, \alpha+\epsilon]  \right\} = f(\alpha).
\label{eq:LD-proba}
\end{equation}
The mapping $\alpha\mapsto f(\alpha)$ is called the large deviations spectrum (or the rate function). For a given function $\Phi$, it is possible to compute the theoretical large deviations spectrum from the rate matrix $A$ as follows. One first computes, for each values of $q\in\R$, the quantity $\Lambda (q)$ defined as the principal eigenvalue (\emph{i.e.,} the largest) of the matrix with elements $A_{ij}+q \delta_{ij}\Phi(j)$ ($\delta_{ij}=1$ if $i=j$ and 0 otherwise). Then the large deviations spectrum can be computed as the Legendre transform of $\Lambda$: 
\begin{equation}
f(\alpha) = \sup_{q\in\R} \left\{q \alpha - \Lambda (q) \right\}, \forall \alpha\in\R.
\label{eq:sup}
\end{equation}

As described in Equation(\ref{eq:LD-proba}), $\alpha_{\tau}=\langle i \rangle_{\tau}$  corresponds in our study case, to the mean number of users $i$ observable over a period of time of length $\tau$ and $f(\alpha)$ relates to the probability of its occurrence as follows:
\begin{equation}
\mathbb{P} \{\langle i \rangle_{\tau} \approx \alpha \} \sim e^{\tau \cdot f(\alpha)}.
\label{eq:joint-prob}
\end{equation}

Interestingly also, if the process is strictly stationary (\emph{i.e.} the initial distribution is invariant) the same large deviation spectrum $f(\cdot)$ can be estimated from a single trace, provided that it is "long enough'' \cite{Barral11a}. We proceed as follows: At a scale $\tau$, the trace is chopped into $k_{\tau}$ intervals $\{I_{j,\tau}=[(j-1)\tau, j\tau[,\,j=1,\ldots,k_{\tau}\}$ of length $\tau$ and we have (almost-surely), for all $\alpha\in\R$:
\begin{equation}
\hspace*{-7mm}
\begin{array}{c}
\displaystyle{f_{\tau}(\alpha, \epsilon_{\tau}) = \frac{1}{\tau} \log \frac{ \#\left\{ j:   \int_{I_{j,\tau}} \Phi(X_s) \ud s \in [\alpha - \epsilon_{\tau}, \alpha+\epsilon_{\tau}]  \right\} }{k_{\tau}} } \\[4mm]
 \mbox{and }\displaystyle{ \lim_{\tau\to\infty} f_{\tau}(\alpha, \epsilon_{\tau}) =  f(\alpha).}
 \end{array}
\label{eq:falpha}
\end{equation}


In practice, for the empirical estimation of the large deviations spectrum, we use a similar  estimator as the one derived in \cite{Barral11b} and also used in \cite{Loiseau10a}.  At scale $\tau$, we compute for each $q\in \R$ the values of $\Lambda_{\tau}^{\prime} (q)$ and $\Lambda_{\tau}^{\prime\prime} (q)$, where 
{$\Lambda_{\tau} (q) = \tau^{-1} \log \left(k_{\tau}^{-1}\sum_{j=1}^{k_t} \exp \left( q \int_{I_{j,\tau}}  \Phi(X_s) \ud s  \right)\right)$}. Then, for each value of $\tau$, we count the number of intervals $I_{j, \tau}$ verifying the condition in expression (\ref{eq:falpha}) and estimate the scale-dependant empirical {\em log-pdf} $f_{\tau}(\alpha, \epsilon_{\tau})$, with the  adaptive choices derived in \cite{Barral11b}:
\begin{equation}
\alpha_{\tau}  = \Lambda_{\tau}^{\prime} (q) \quad \textrm{ and }  \quad \epsilon_{\tau} = \sqrt{ \frac{-\Lambda_{\tau}^{\prime\prime} (q)}{\tau} }.
\label{eq:alpha}
\end{equation}
Let us now illustrate the LDP in the context of the specific VoD use case, where $X$ would correspond to $(i, r)$, the bi-variate Markov process. $\Phi(X)$ is $i$, the observable and
$\int^\tau_0 \Phi(X_{S})\,ds = \langle i \rangle_{\tau}$ corresponds to the average number of users with a period $\tau$.
\subsection{Numerical Interpretations}
\label{sec:interpretation}
\begin{figure}[t]
\begin{center}
\begin{tabular}{ccc}
{\normalsize (a)} & \hspace*{-8mm} {\normalsize (b)} & \hspace*{-8mm} {\normalsize (c)} \\
\hspace*{-10mm}\includegraphics[width=0.35\columnwidth]{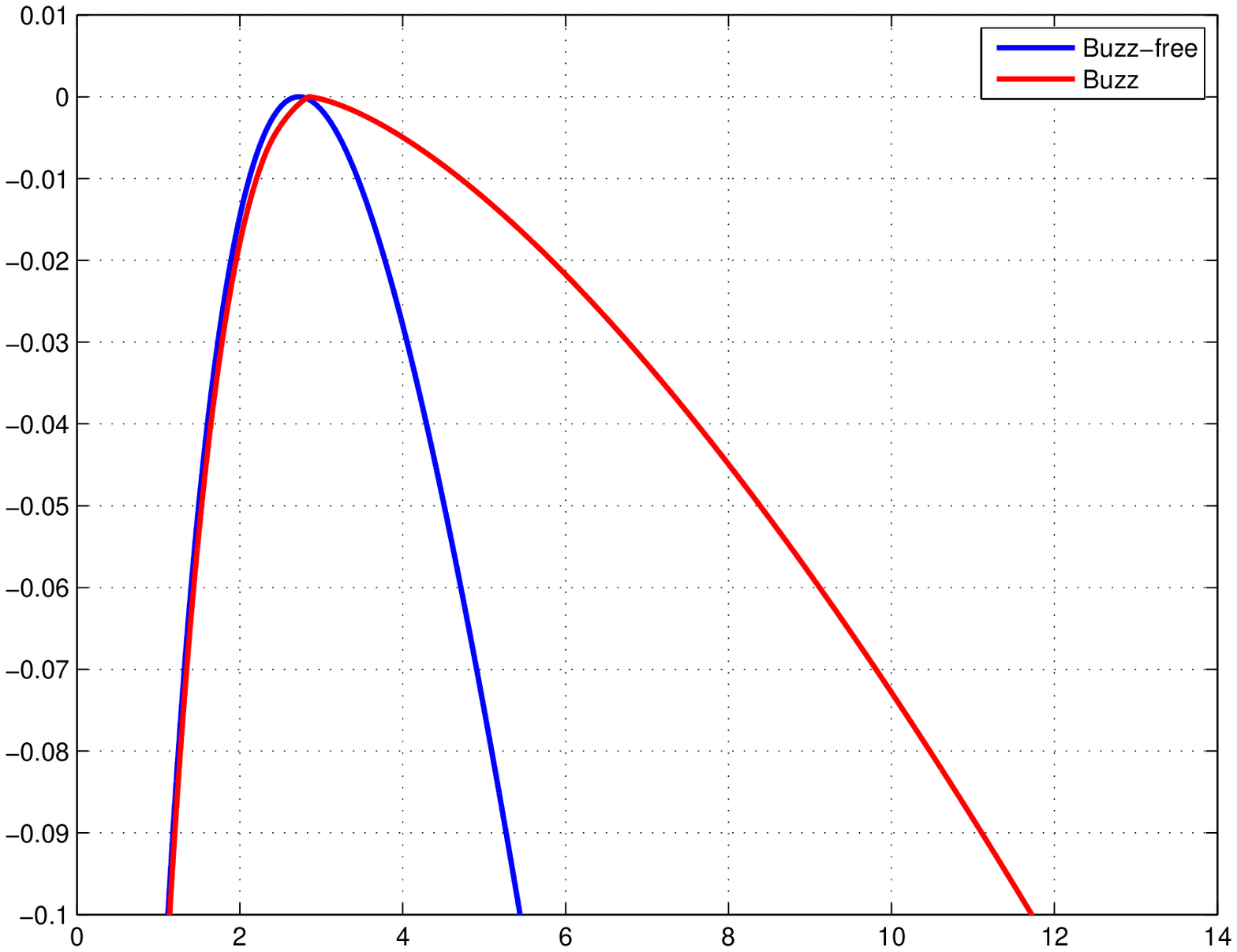}
&
\hspace*{-0mm}\includegraphics[width=0.35\columnwidth]{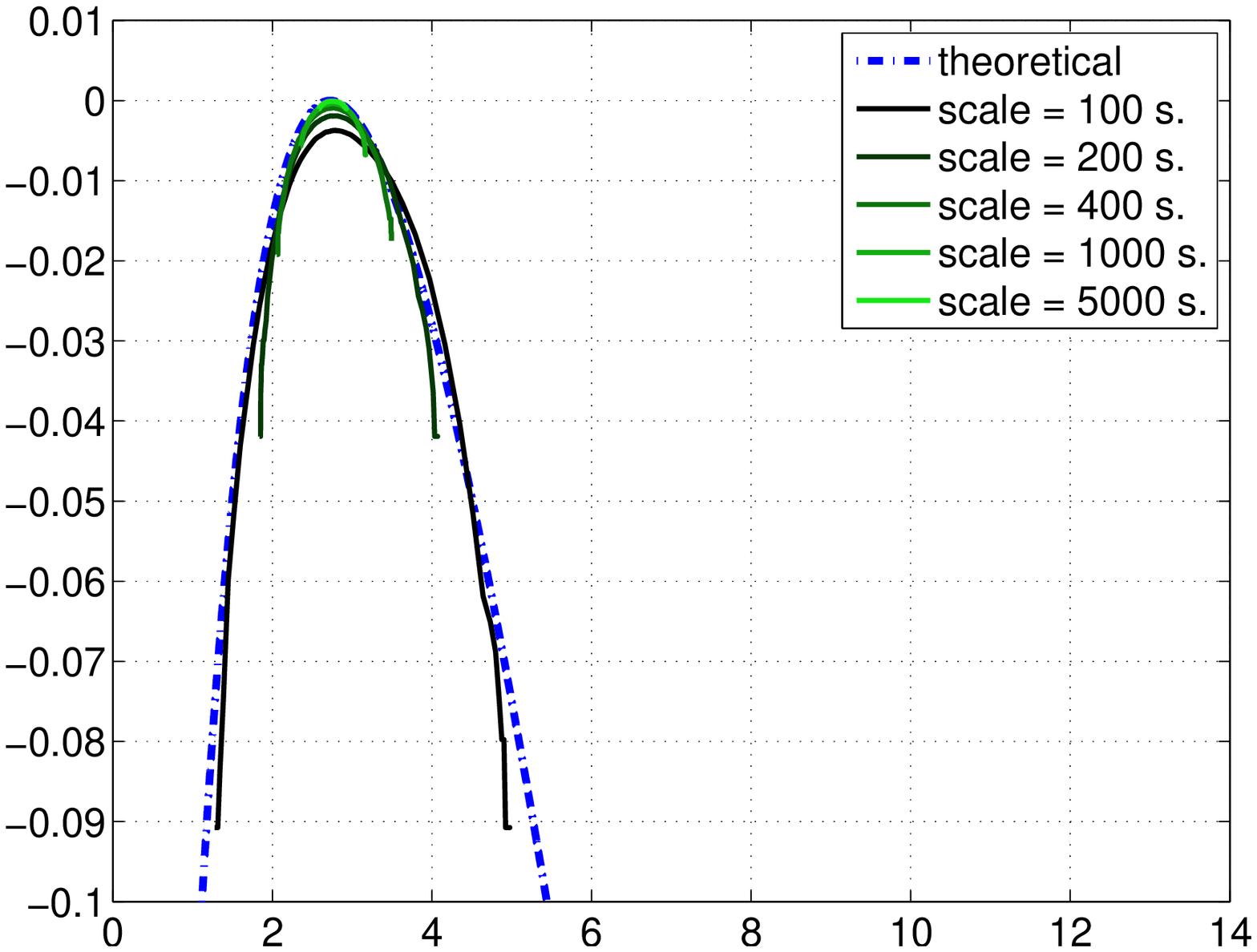} 
&
\hspace*{-0mm}\includegraphics[width=0.35\columnwidth]{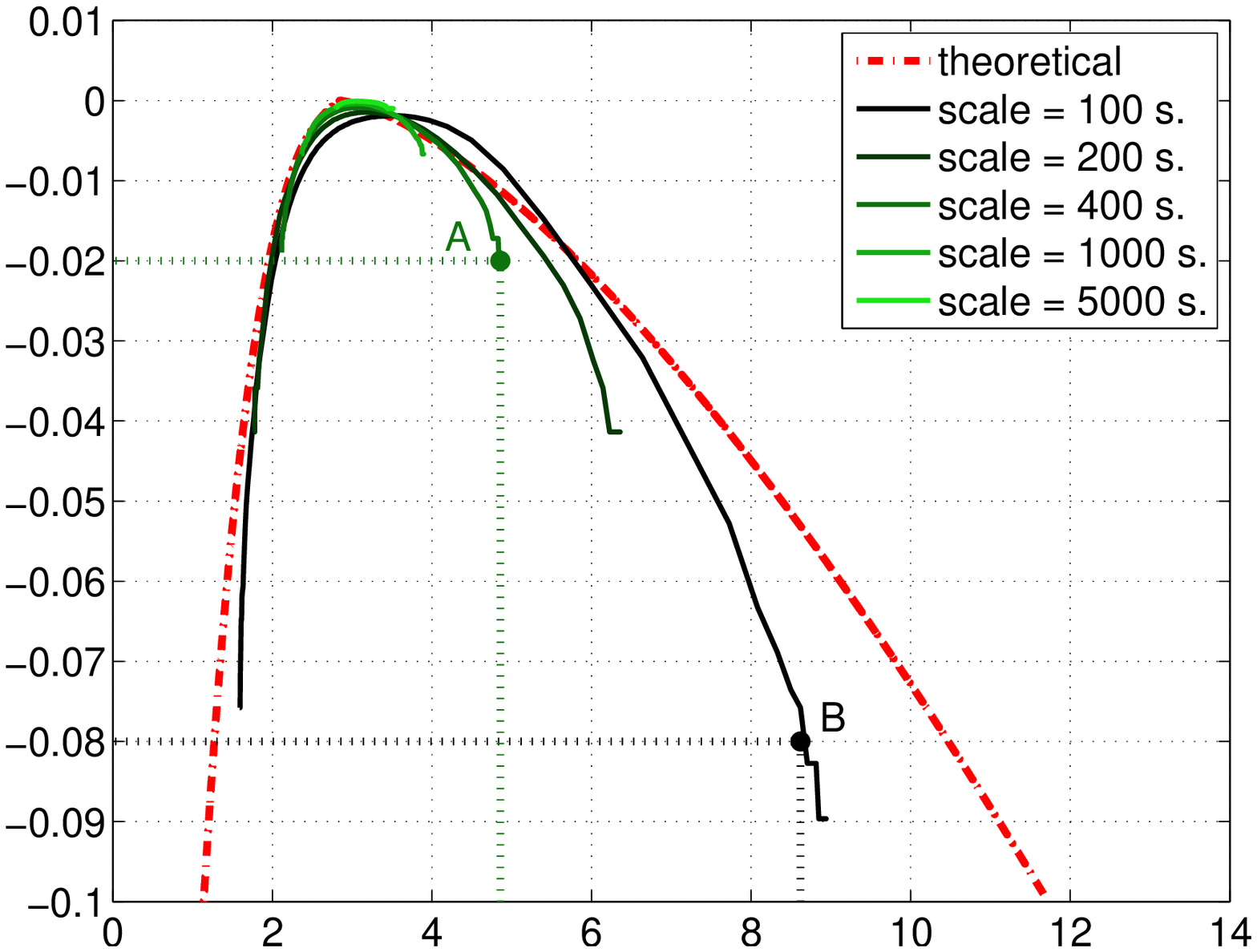}\\[-26mm]
\hspace*{-72mm}\begin{turn}{90}$f(\alpha)$\end{turn} 
& 
\hspace*{-56mm}\begin{turn}{90}$f(\alpha)$\end{turn} 
&
\hspace*{-56mm}\begin{turn}{90}$f(\alpha)$\end{turn}  \\[20mm]
$\alpha = \langle i \rangle_{\tau}$ & $\alpha = \langle i \rangle_{\tau}$ & $\alpha = \langle i \rangle_{\tau}$  \\
\end{tabular}
\end{center}
\caption{\small Large Deviations spectra corresponding to the traces of Figure \ref{fig:traces}. (a) Theoretical spectra for the buzz free (blue) and for the buzz (red) scenarii. (b) \& (c) Empirical estimations of $f(\alpha)$ at different scales from the buzz free and the buzz traces, respectively. }
\label{fig:LDP}
\end{figure}

For ease of computation we estimate the Large Deviation Spectrum for cases where $I_{\rm max} = 30, R_{\rm max}=60$. We also choose the parameters accordingly (so that it does not saturate with the maximum value) for buzz and buzz-free scenarios. For the first case $\beta_{1} = 0.1$, $\beta_{2} = 0.8$, $\gamma = 0.7$, $\mu = 0.3$, $l = 1.0$, $a_{1}=0.006$ and $a_{2}=0.6$. For the buzz-free case: $\beta_1=\beta_2=\beta= 0.1$, $\gamma = 0.7$, $\mu = 0.3$, $l = 1.0$. 
Intrinsically, Large Deviation Principle naturally embeds the time scale notion into the statistical description of the aggregated observable at different time resolutions. As expected, the theoretical LD spectra displayed in Figure~\ref{fig:LDP}(a) reach their maximum for the same mean number of users. This apex is the almost sure value as described in Section~\ref{sec:LDP}. As the name suggests almost sure workload ($\alpha_{a.s}$) corresponds to the mean value that we almost surely observe on the trace. More interestingly though, the LD spectrum corresponding to the buzz case, spans over a much larger interval of observable mean workloads than that of the buzz-free case. This remarkable support widening of the theoretical spectrum shows that LDP can accurately quantify the occurrence of extreme, yet rare events.

Plots (b)-(c) of Figure \ref{fig:LDP}  compare theoretical and empirical large deviation spectra obtained for the two traces. For each given scale ($\tau$) the empirical estimation procedure yields one LD estimate. These empirical estimates at different scales superimpose for a given range of $\alpha$. This is reminiscent of the scale invariant property underlying the large deviation principle. If we focus on the supports of the different estimated spectra, the larger the time scale $\tau$ is, the smaller becomes the interval of observable value of $\alpha$. This is coherent with the fact that for a finite trace-length the probability to observe a number of current viewers, that in average, deviates from the nominal value ($\alpha_{a.s}$) during a period of time ($\tau$) decreases exponentially fast with $\tau$. To fix the ideas, the estimates of plot (c),  indicate that  for a time scale $\tau =400\,sec.$, the maximum observable mean number of users is around 5 with probability $2^{400\cdot(-0.02)} \approx 35.10^{-5}$ (point A), while it increases up to $9$ with the same probability ($2^{100\cdot(-0.08)}$) for $\tau =100\,sec.$ (point B).

\newpage

\section{Resource management policies}
\label{sec:management}
Retuning to our VoD use case, we now sketch two possible schemes for exploiting the Large Deviation description of the system to dynamically provision the allocated resources:

\begin{itemize}
\item {\it Identification of the reactive time scale for reconfiguration}: Find a relevant time scale that realizes a good trade-off between the expectable level of overflow associated to this scale and a sustainable {\sc opex} cost induced by the resources reconfiguration needed to cope with the corresponding flash crowd.
\item {\it Link capacity dimensioning}: Considering a maximum admissible loss probability, find the  safety margin  that it is necessary to provision on the link capacity, to guarantee the corresponding QoS. 
\end{itemize}

\subsection{Identification of the reactive time scale for reconfiguration}
\label{sec:scale}

We consider the case of a VoD service provider who wants to determine the reactivity scale at which it needs to reconfigure its resource allocation. This quantity should clearly derive from a good compromise between the level of congestion (or losses) it is ready to undergo, i.e. a tolerable performance  degradation, and the price it is willing to pay for a frequent reconfiguration of its infrastructure. Let us then assume that the VoD provider has fixed admissible bounds for these two competing factors, having determined the following quantities:
\begin{itemize}
\item $\alpha^* > \alpha_{\rm a.s.}$: the deviation threshold  beyond which it becomes worth (or mandatory) considering to reconfigure the resource allocation. This choice is uniquely determined by a {\sc capex} performance concern. 
\item $\sigma^*$: an acceptable probability of occurrence of these overflows. This choice is essentially guided by the corresponding {\sc opex} cost.
\end{itemize}

Let us moreover suppose, that the LD spectrum $f(\alpha)$ of the workload process was previously estimated, either by identifying the parameters of the Markov model used to describe the application, or empirically from collected traces. Then, recalling the probabilistic interpretation we surmised in relation (\ref{eq:joint-prob}), the minimum reconfiguration time scale $\tau^*$ for dynamic resource allocation, that verifies the sought compromise, is simply the solution of the following inequality:
\begin{equation}
\tau^* = \max\{\tau : \mathbb{P}{\{ \langle i \rangle_{\tau} \geq \alpha^* \} }  =  \int^\infty_{\alpha^*} e^{\tau f_{\tau}(\alpha)}\,d{\alpha}  \geq \sigma^*\},
\label{eq:meanusr}
\end{equation}
with $f_{\tau}(\alpha)$ as defined in expression (\ref{eq:falpha}).

From a more general perspective though, we can see this problem as an underdetermined system involving 3 unknowns ($\alpha^*$,$ \tau^*$ and $\sigma^*$) and only one relation (\ref{eq:meanusr}). Therefore, and depending on the sought objectives, we can imagine to fix any other two of these variables and to determine the resulting third so that it abides with the same inequality as in expression (\ref{eq:meanusr}).

\subsection{Link capacity dimensioning}
\label{sec:serv}

We now consider an architecture dimensioning problem from  the infrastructure provider perspective. Let us assume that the infrastructure and the service providers have come to a Service Level Agreement (SLA), which among  other things, fixes a tolerable level of losses due to link congestion.  We start considering the case of a single VoD server and address the following question: What is the minimum link capacity $C$ that has to be provisioned  such that we meet the negotiated QoS in terms of loss probability? Like in the previous case, we assume that the estimated LD spectrum $f(\alpha)$ characterizing the application has been priorly identified. A rudimentary SLA would be to  guarantee a loss free transmission for  the {\em normal} traffic load only: this loose QoS would simply amount to fix $C$ to the almost sure workload $\alpha_{\rm a.s.}$. Naturally then, any load overflow beyond this value will result in goodput limitation (or losses, if there is no buffer to smooth out exceeding loads). For a more demanding QoS, we are led to determine the necessary safety margin $C_0>0$ one has to provision above $\alpha_{\rm a.s.}$ to absorb the exact amount of overruns corresponding to the loss probability $p_{\rm loss}$ that was negotiated in the SLA.  From the interpretation of the large deviation spectrum provided in Section \ref{sec:LDP},  this margin $C_0$ is determined by the resolution of the following inequality:
\begin{eqnarray}
C_0 & \:{\rm :} &  \int_{\alpha_{\rm a.s.}+C_0}^{\infty} \int_{\tau_{min}}^{\tau_{max}} e^{\tau\cdot f(\alpha)}\,{\rm d}\tau\,{\rm d}\alpha~ \leq ~p_{\rm loss} \nonumber \\\label{eq:ploss-1}
& \:{\rm :} &\int_{\alpha_{\rm a.s.}+C_0}^{\infty} \frac{e^{\tau_{max}\cdot f(\alpha)}-e^{\tau_{min}\cdot f(\alpha)}}{f(\alpha)} \,{\rm d}\alpha ~\leq ~p_{\rm loss}
\end{eqnarray}
In this expression,  $\tau_{min}$ is typically  determined by the size $Q$ of the buffers that is usually provisioned to dampen the traffic volatility. In that case,
\begin{equation}
\tau_{\rm min} = \frac{Q}{\alpha-(\alpha_{a.s.}+C_0)},
\label{eq:tau_min}
\end{equation} 
corresponds to the maximum burst duration that can be buffered without causing any loss at rate $\alpha>C=\alpha_{a.s.}+C_0$.  
As for $\tau_{\rm max}$, it relates to the maximum period  of reservation dedicated to the application. Most often though, the characteristic time scale of the application exceeds the dynamic scale of flash crowds  by several orders of magnitude, and $\tau_{\rm max}$ can then simply be set to infinity. With these particular integration bounds, Equation (\ref{eq:ploss-1}) simplifies to
\begin{equation}
\begin{array}{c}
\displaystyle{C_0 = C-\alpha_{a.s.} \:{\rm :} \int_{C}^{\infty} \frac{-1} {f(\alpha)}\,e^{\frac{Q}{\alpha-C} f(\alpha)}\,{\rm d}\alpha ~ \leq ~ p_{\rm loss}},
\\[5mm]
\end{array}
\label{eq:ploss-2}
\end{equation}
a decreasing function of $C$, which can be solved using a simple bisection technique. \\
As long as the server workload remains below $C$,  this resource dimensioning  guarantees that no loss occurs. All overrun above this value will produce losses, but we ensure that the frequency (probability) and duration of these overruns are such that the loss rate remains  conformed to the SLA. 
The proposed approach clearly contrasts  with resource over-provisioning  that does not seek at optimizing the {\sc capex} to comply with the loss probability tolerated in the SLA.

The same provisioning scheme can straightforwardly be generalized to the case of several applications sharing a common set of resources. To fix the idea, let us consider an infrastructure provider that wants to  host $K$ VoD servers over the same shared link. A corollary question is then to determine how many servers $K$ can the fixed link capacity $C$ support, while guaranteeing  a prescribed level of losses. If the servers are independent, the probability for two of them to undergo a flash crowd simultaneously is negligible. For ease and without loss of generality, we moreover suppose that they are identically distributed and modeled by the same LD spectrum $f^{(k)}(\alpha)=f(\alpha)$ with the same nominal workload $\alpha^{(k)}_{\rm a.s.}=\alpha_{\rm a.s.},\,k=1,\ldots K$. 
Then, following the same reasoning as in the previous case of a single server, the maximum number $K$ of servers reads:

\begin{equation}
K = \mbox{arg}\max_{K} \left( C -K\cdot \alpha_{\rm a.s.}\right) \leq C_0,
\end{equation}
where the safety margin $C_0$ is defined as in expression (\ref{eq:ploss-2}).

Then, depending on the agreed {\it Service Level Agreements}, the infrastructure provider can easily offer different levels of probability losses (QoS) to its VoD clients, and adapt  the number of hosted servers, accordingly.
\begin{figure} [h]
\centering
\begin{tabular}{c}
\includegraphics[width=0.4\columnwidth]{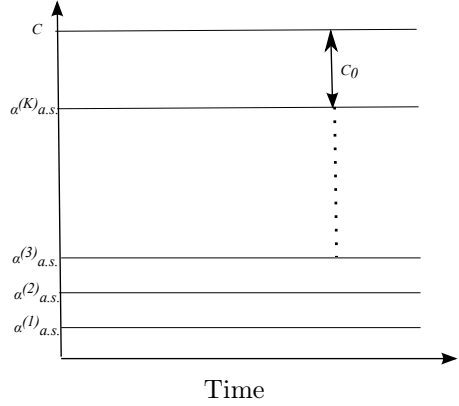}\\
{\normalsize Time}
\end{tabular}
\caption{\small Dimensioning $K$, the number of hosted servers sharing a fixed capacity link $C$. The safety margin $C_0$ is determined according to the probabilistic loss rate negotiated in the {\it Service Level Agreement} between the infrastructure provider and the VoD service provider.}
\label{fig:caplan}
\end{figure}
\newpage

\section{Conclusion}
\label{sec:conclusion}
Many applications deployed on a cloud infrastructure, such as a Video on Demand service, are well known for undergoing highly volatile demand, making their workload hard to qualitatively and quantitatively characterize.
Adopting a constructive approach to capture the VoD users' behavior, in this report we  proposed a simple, concise and versatile model for generating the workload variations in such context.
We also devised an heuristic identification procedure that aims at estimating the parameters values of the model from a single collected trace. 
First, we numerically evaluated the accuracy of this procedure using several synthetic traces. 
Our experiments show that the procedure introduces little bias and typically recovers the actual parameters value with a relative error of about 10\%. 
Second, we apply this same procedure against two real video server workload traces. 
Obtained results demonstrate that, once the model has been calibrated, it succeeds to reproduce the statistical behavior of the real trace (in terms of both the steady-state probabilities and the autocorrelations for the workload time series).
Moreover, owing to the constructive nature of our model, the estimated values of the parameters provide valuable insight on the application that it would be difficult, or even impossible, to infer  from the raw traces. The captured information may answer questions of practical interest to cloud oriented service providers, like: is the application workload mostly driven by spontaneous behaviors, or is it rather subject to a  gossip phenomenon?

Furthermore, a key-point of this model is that it permits to reproduce the workload time series with a Markovian process, which is known to verify a Large Deviation Principle (LDP). This particularly interesting property yields a large deviation spectrum whose interpretation enriches the information conveyed by the standard steady state distribution: For a given observation (workload trace), LDP allows to infer (theoretically and empirically) the probability that the time average workload, calculated at an arbitrary aggregation scale, deviates from its nominal value (i.e. almost sure value). 

We leveraged this multiresolution probabilistic description to conceptualize two different management schemes for dynamic resource provisioning. As explained, the rationale is to use large deviation information to help network and service providers together to agree on the best {\sc capex}-{\sc opex} trade-off. Two major stakes of this negotiation are: {\it (i)} to determine the largest reconfiguration time scale adapted to the workload elasticity and {\it (ii)} to dimension VoD server so as to guarantee with upmost probability the Quality of Service imposed by the negotiated  Service Level Agreement.
\newline
More generally though, the same LDP based concepts can benefit any other ``Service on Demand" scenarii to be deployed on dynamic cloud environments.  \newpage

\bibliographystyle{IEEEbib}
\bibliography{rr_reso}

\end{document}